\documentclass[aps, prd, 12pt, eqsecnum, tightenlines, notitlepage, superscriptaddress, nofootinbib, floatfix]{revtex4-2}
 
\usepackage{aas_macros}
\usepackage{amsfonts}
\usepackage[utf8]{inputenc}
\usepackage[margin=1in]{geometry}
\usepackage{graphicx}
\usepackage{amsmath}
\usepackage{caption, subcaption}
\usepackage{hyperref}
 \usepackage{makecell}
 \usepackage{float}
 \usepackage[usenames, dvipsnames]{xcolor}
\usepackage{ulem}
\usepackage{soul,xcolor}
\hypersetup{colorlinks,bookmarksnumbered,
linkcolor=Corange,citecolor=Corange,urlcolor=Corange}

\setstcolor{red}

\usepackage{color, colortbl}
\definecolor{Corange}{RGB}{255,108,12}
\definecolor{CompCorange}{RGB}{0,59,76}
\definecolor{Gray}{gray}{0.8}
\definecolor{GrayLight}{gray}{0.4}
\definecolor{Darkgreen}{RGB}{30,120,30}
\definecolor{granate}{rgb}{0.8039,0.2,0.2}
\definecolor{krisscolor}{RGB}{3, 156, 11}

\newcommand{\e}[1]{\times 10^{#1}}

\usepackage{natbib}

\newcommand{\beq}{\begin{equation}}
\newcommand{\eeq}{\end{equation}}
\newcommand{\bea}{\begin{eqnarray}}
\newcommand{\eea}{\end{eqnarray}}

\usepackage{tikz}
\usetikzlibrary{"arrows", "automata", "backgrounds", "calendar", "chains", "matrix", "mindmap", "patterns", "petri", "shadows", "shapes.geometric", "shapes.misc", "spy", "trees"}
\usetikzlibrary{arrows,shapes}
\usetikzlibrary{trees}
\usetikzlibrary{matrix,arrows} 				% For commutative diagram
\usetikzlibrary{positioning}				% For "above of=" commands
\usetikzlibrary{calc,through}				% For coordinates
\usetikzlibrary{decorations.pathreplacing}  % For curly braces
\usepackage{pgffor}							% For repeating patterns

\usetikzlibrary{decorations.pathmorphing}	% For Feynman Diagrams
\usetikzlibrary{decorations.markings}
\tikzset{
    vector/.style={decorate, decoration={snake}, draw},
	provector/.style={decorate, decoration={snake,amplitude=2.5pt}, draw},
	antivector/.style={decorate, decoration={snake,amplitude=-2.5pt}, draw},
    fermion/.style={draw=black, postaction={decorate},
        decoration={markings,mark=at position .55 with {\arrow[draw=black]{>}}}},
    fermioncyan/.style={draw=black, postaction={decorate},
        decoration={markings,mark=at position .55 with {\arrow[draw=cyan]{<}}}},
    fermiondif/.style={draw=black, postaction={decorate},
        decoration={markings,mark=at position .7 with {\arrow[draw=black]{>}}}},
            fermiondif2/.style={draw=black, postaction={decorate},
        decoration={markings,mark=at position .7 with {\arrow[draw=black]{<}}}},
    fermionend/.style={draw=black, postaction={decorate},
        decoration={markings,mark=at position 1 with {\arrow[draw=black]{>}}}},
    fermionuchannel2/.style={draw=black, postaction={decorate},
        decoration={markings,mark=at position .4 with {\arrow[draw=black]{>}}}},
    scalardif/.style={dashed,draw=black, postaction={decorate},
        decoration={markings,mark=at position .7 with {\arrow[draw=black]{>}}}},
    scalarend/.style={dashed,draw=black, postaction={decorate},
        decoration={markings,mark=at position 1 with {\arrow[draw=black]{>}}}},
    fermionbar/.style={draw=black, postaction={decorate},
        decoration={markings,mark=at position .55 with {\arrow[draw=black]{<}}}},
    fermionnoarrow/.style={draw=black},
    gluon/.style={decorate, draw=black,
        decoration={coil,amplitude=4pt, segment length=5pt}},
    scalar/.style={dashed,draw=black, postaction={decorate},
        decoration={markings,mark=at position .55 with {\arrow[draw=black]{>}}}},
    scalarcyan/.style={dashed,draw=black, postaction={decorate},
        decoration={markings,mark=at position .55 with {\arrow[draw=cyan]{>}}}},
    scalaruchannel1/.style={dashed,draw=black, postaction={decorate},
        decoration={markings,mark=at position .7 with {\arrow[draw=black]{>}}}},
                  scalaruchannel2/.style={dashed,draw=black, postaction={decorate},
        decoration={markings,mark=at position .4 with {\arrow[draw=black]{>}}}},
    scalarbar/.style={dashed,draw=black, postaction={decorate},
        decoration={markings,mark=at position .55 with {\arrow[draw=black]{<}}}},
    scalarnoarrow/.style={dashed,draw=black},
    electron/.style={draw=black, postaction={decorate},
        decoration={markings,mark=at position .55 with {\arrow[draw=black]{>}}}},
	bigvector/.style={decorate, decoration={snake,amplitude=4pt}, draw},
}

\tikzstyle{block} = [draw, rectangle, 
    minimum height=3em, minimum width=6em]

\usepackage{tikz}
\usetikzlibrary{fit}
\tikzset{%
  highlight/.style={rectangle,rounded corners,color=granate,draw,text opacity =1,
    fill opacity=0.5,thick,inner sep=0pt}
}

\usepackage{placeins}

\tikzset{
    cross/.pic = {
    \draw[rotate = 45] (-#1,0) -- (#1,0);
    \draw[rotate = 45] (0,-#1) -- (0, #1);
    }
}

\tikzset{
    square/.style={%
        draw=none,
        circle,
        append after command={%
            \pgfextra \draw[#1] (\tikzlastnode.north-|\tikzlastnode.west) rectangle 
                (\tikzlastnode.south-|\tikzlastnode.east);\endpgfextra}
    },
    square/.default=black
}

\tikzstyle{block} = [draw, rectangle, 
    minimum height=3em, minimum width=6em]

\usepackage{xparse}
\NewDocumentCommand\semiloop{O{black}mmmO{}O{above}}
{%
\draw[#1] let \p1 = ($(#3)-(#2)$) in (#3) arc (#4:({#4+180}):({0.5*veclen(\x1,\y1)})node[midway, #6] {#5};)
}

\begin{document}
\hspace{2.2in} \mbox{CALT-TH/2023-030}

\preprint{APS/123-QED}

\title{Contrast Loss from Astrophysical Backgrounds in Space-Based Matter-Wave Interferometers}
\author{Yufeng Du}
\affiliation{Walter Burke Institute for Theoretical Physics, California Institute of Technology, Pasadena, CA 91125, USA}
\author{Clara Murgui}
\affiliation{Walter Burke Institute for Theoretical Physics, California Institute of Technology, Pasadena, CA 91125, USA}
\author{Kris Pardo}
\affiliation{Department of Physics \& Astronomy, University of Southern
California, Los Angeles, CA, 90089, USA}
\affiliation{Walter Burke Institute for Theoretical Physics, California Institute of Technology, Pasadena, CA 91125, USA}
\author{Yikun Wang}
\affiliation{Walter Burke Institute for Theoretical Physics, California Institute of Technology, Pasadena, CA 91125, USA}
\author{Kathryn M. Zurek}
\affiliation{Walter Burke Institute for Theoretical Physics, California Institute of Technology, Pasadena, CA 91125, USA}
\date{\today}
\begin{abstract}
Atom and matter interferometers are precise quantum sensing experiments that can probe differential forces along separated spacetime paths. Various atom and matter interferometer experiments have been proposed to study dark matter, gravitational waves, and exotic new physics. Increasingly, these experimental concepts have proposed space-based designs to maximize interrogation times and baselines. However, decoherence and phase shifts caused by astrophysical backgrounds could largely undermine or destroy the target sensitivity of the experiments. We calculate the decoherence effects induced by solar photons, the solar wind, cosmic rays, solar neutrinos and zodiacal dust on space-based atom and matter interferometers. We find that, in future space-based atom and matter interferometers, the solar wind generically produces decoherence beyond the quantum noise limit, without proper shielding. In addition, solar photons are also an important background for matter interferometers. 
\end{abstract}
\maketitle

\vfill

\tableofcontents

\pagebreak

%%%%%%%%%%%%%%%%%%%%%%%%
\section{Introduction}
%%%%%%%%%%%%%%%%%%%%%%%%
%
Atom and matter interferometers are becoming increasingly precise quantum sensing experiments.  Atom interferometers use coherent atomic wavepackets to measure the differential forces along two ``arms'' (for a general review of atom interferometers, see Ref.~\cite{Geiger2020}). For concreteness, consider a Mach-Zehnder atom interferometer \citep{Kasevich1991}. Analogously to the conventional Mach-Zehnder interferometer, here the de-Broglie wave nature of the atoms plays the role of light, and laser pulses act as the beamsplitters and mirrors. The atoms start in a coherent wavepacket, then are split by a laser pulse into two different wavepackets, which travel along two different paths separated by macroscopic distances. Two more laser pulses reverse the motions of these wavepackets and recombine them. The state of the recombined wavepacket is then read, with the relative populations depending on the initially prepared state, the timing of the laser pulses, and any differential forces felt along the two paths. Matter interferometers work similarly, but the spatial superpositions occur at the level of the material object. Matter-wave interferometers\footnote{We use ``matter-wave interferometers'' to refer to both atom and matter interferometers.} have proven to be powerful tests of the equivalence principle, {\it e.g.}~\cite{Asenbaum2020}, and promising probes of gravitational waves and dark matter \citep{Dimopoulos:2008sv, Arvanitaki2015, Graham2016, Arvanitaki:2016fyj, Geraci2016, Badurina:2021rgt}. 

The current and next generation stages of matter-wave interferometers are all terrestrial \citep{Asenbaum2020, Overstreet2022, MAGIS-100:2021etm}. Because the sensitivity of matter-wave interferometers for most of their science objectives is set by their phase sensitivity, matter-wave interferometers seek long interrogation times. This has led to tall towers on Earth, or deep mine shafts. However, the best interrogation times and longest baselines can only be achieved in microgravity. This reality has led to many proposals for space-based matter-wave interferometers, {\it e.g.} \cite{Frye2021, Kaltenbaek2016, AEDGE, GDM}. Although many of these proposals are futuristic, their promise as powerful and multi-purpose scientific instruments warrants their careful consideration.

An important assumption that has been made in all space-based experimental proposals is that they can be properly shielded from any particle backgrounds. Scattering of particles on the atomic clouds (or solid objects) in these experiments risks decohering the system, becoming potential backgrounds in new physics searches. For matter interferometers, some backgrounds could decohere the system to a degree precluding a fringe measurement. In the solar system, these experiments would be buffeted by solar photons, solar wind particles, cosmic rays, solar neutrinos, and zodiacal dust. There are terrestrial, experimental constraints on the rate of decoherence from gas particles \citep{Hornberger2003b}, as well as a long literature regarding theoretical calculations of these decoherence rates (see, {\it e.g.},~\cite{Joos:1984uk, Gallis1990, Hornberger_2003, Adler_2006}). In addition, recent work has computed the decoherence from long-range forces \citep{Kunjummen:2022uzx}. However, there are no detailed, particle-based calculations of the rates of these space-based backgrounds. 

In this paper, we derive the decoherence effects of various particle backgrounds on space-based matter-wave interferometers. We show that some backgrounds, like solar photons and the solar wind, could prove to be important without proper shielding. In Section~\ref{sec:overview}, we provide an overview of the different backgrounds and experiments that we consider in this paper, and we give a brief summary of our results. In Section~\ref{sec:formalism}, we introduce our theoretical formalism. In Sections~\ref{sec:photons} \& \ref{sec:crs}, we discuss the most important backgrounds: photons and charged particles. In Section~\ref{sec:other_bkgds}, we consider other particle backgrounds. We review our results and conclude in Section~\ref{sec:conclusion}.

\section{Overview and Summary of Results}\label{sec:overview}

In this work, we consider the decoherence effects of background particles on matter-wave interferometers in space. In this section, we first describe the observable effects that these particles produce (\ref{sec:overview_effects}). We then give an overview of the four main backgrounds we consider: solar photons, charged particles ({\it i.e.}, the solar wind and galactic cosmic rays), zodiacal dust particles, and solar neutrinos. We discuss the typical energies and fluxes, and provide a summary of their decoherence effects on matter-wave interferometers (\ref{sec:overview_backrounds}). A full discussion of each process is relegated to Sections~\ref{sec:photons}, \ref{sec:crs} \& \ref{sec:other_bkgds}. We briefly describe the various matter-wave interferometer experiments we consider in this paper (\ref{sec:exps}).  Finally, in Section~(\ref{sec:summary}), we summarize our results.

\subsection{Observables in Matter-Wave Interferometers}\label{sec:overview_effects} 

First, let us discuss the main observable effects that background particles would induce on space-based matter-wave interferometers. The two ``arms'' of matter-wave interferometers probe the spacetime and differential forces along two paths. Matter-wave interferometers begin with a coherent state, which is then split such that wavepackets in an excited state separate from the ground state wavepackets and follow physically separated paths. When the wavepackets are recombined, the amount of decoherence and any coherent phase shifts induced by environmental effects can be measured. Particle scattering can lead to both decoherence and phase shifts in the system \citep{Joos:1984uk, Hornberger_2003, Riedel:2012ur, Riedel2017, AIDM}. Decoherence measurements can be used to measure particle dark matter scattering \cite{AIDM}, and quantum gravity signatures \cite{Pino2016}. The phase is the relevant observable for gravity, gravitational wave, and ultralight dark matter measurements \citep{Asenbaum2020, Dimopoulos:2008sv, Arvanitaki2015, Graham2016, Arvanitaki:2016fyj, Geraci2016, Badurina:2021rgt}. However, if any background decoherence effects are sufficiently robust, the fringe would be destroyed and phase shifts could not be measured. In this paper, we focus on the decoherence effects only and leave the phase effects to future work.

To quantify the amount of decoherence, we must discuss the actual observables further. For concreteness, consider an unentangled atom interferometer (\textit{e.g.,} a cold atom interferometer) 
with $N_{\rm atoms}$ independent atoms. In what follows, we will use $N_{\rm{ind}}$ to mean the number of {\it independent} objects that are measured at the read-out port. So, a cold atom interferometer has $N_{\rm{ind}} = N_{\rm{atoms}}$. The density matrix parametrizing the two-level system of each of these independent objects (atoms in a cold atom interferometer) is:
\begin{equation}\label{eq:densitymat}
   \rho = \frac{1}{2} \left(\begin{matrix} 1 & \gamma \\
   \gamma^* & 1\end{matrix}\right) \; ,
\end{equation}
where $\gamma = \exp (-s+i\phi)$ is the decoherence factor, which can be decomposed into a dimensionless decoherence, $s$ (real, positive), and a phase, $\phi$ (real). 
Eq.~\eqref{eq:densitymat} is displayed in the basis of the state in which the object is initially prepared and the higher energy level that gives the macroscopic wavepacket separation $\Delta x$. 

At the end of the interrogation time, the two wavepackets are recombined and the atom states are read. Atom interferometer experiments measure the relative populations of atoms in the final, recombined wavepackets~\cite{Geiger2020,Roura:2014vwa}\footnote{
Most modern matter-wave interferometers produce a fringe measurement, similar to those produced in light interferometers, that allows them to measure both the visibility and phase of each measurement. This can be achieved either via the cloud preparation, modifications to the detection scheme, or through the use of multiple matter-wave interferometers \citep{Dickerson2013, Sugarbaker2013, Foster:02, Chiow2009}. However, we find the original measurement scheme simpler to use for our explanations here.}. The measurement is performed in the states defining the basis of Eq.~\eqref{eq:densitymat}: the state in which they were initially prepared, which we call port I (or the bright port), and in the higher energy level state, which we call port II (or the dim port). The probability of measuring an atom in port I is given by: 
\begin{equation}\label{eq:pin}
    p_I = \frac{1}{2}\left( 1 + {\cal R}{\rm e}\{ \gamma \} \right) = \frac{1}{2}\left( 1 + e^{-s}\cos \phi \right) \; .
\end{equation}
In the absence of decoherence ($s=0$), each atom is a pure quantum state and, with the proper final phase $(\phi = 0)$, has a 100\% probability of being measured in port I. Therefore, counts in port II suggest that the system experienced decoherence. For an ensemble of $N_{\rm ind}$ measurements, the number of expected counts in port I, $N_I$, corresponds to the expectation value of the combined probability distribution of the individual atoms in Eq.~\eqref{eq:pin}. This is given by the trace of the 1-body density matrix (after tracing out $N-1$ atoms) with the measurement operator projecting each atom~\cite{Badurina:2024nge}. Assuming all of the atoms can be treated independently (ignoring cooperative effects, in the limit of weak decoherence effects), the combined probability is given by the binomial distribution~\citep{Itano1993},
\begin{equation}\label{eq:binomial}
   P_{I} = \begin{pmatrix} N_{\rm{ind}} \\ N_{I} \end{pmatrix} p_I^{N_I} (1-p_{I})^{N_{II}} \; ,
\end{equation}
where $N_{II} = N_{\rm ind} - N_{I}$. Its expectation value and variance are given by:
\begin{equation} \label{eq:exvar}
\langle P_{I} \rangle = (N_I + N_{II}) \, p_I = N_{\rm ind} \, \, p_I, \qquad  \sigma^2 = N_{\rm ind} \, \, p_I (1-p_I) \; .
\end{equation}
In order to claim that the system has decohered, one needs to compare the number of atoms measured in port II, $N_{II}$, with the variance of the distribution (obtained by marginalizing the probability $p_I$ over $\phi$ with $s=0$), $\bar \sigma =  \sqrt{N_{\rm ind}}/2$, which can be identified as the one-sided shot noise in the large $N_{\rm ind}$ limit.\footnote{The binomial distribution and its variance apply to classical independent measurements of independent states. Ref.~\citep{Itano1993} shows that the same results can be derived for a $2^{N_{\rm ind}}$-dimensional Hilbert space of $N_{\rm ind}$ atoms, constructed from the direct product states of the form: 
\begin{equation}
    |\Psi_1, \Psi_2, \cdots, \Psi_{N_{\rm in}}\rangle = \prod_{\otimes}^{i = 1,\cdots,N_{\rm ind}} |\Psi_i\rangle \; , \nonumber
\end{equation}
where $|\Psi_i\rangle$ are the individual atom states.}

The atom interferometer experiment literature parameterizes the impact of the decoherence on the relative number counts as follows:
\begin{equation}\label{eq:V}
   \frac{N_I}{N_I + N_{II}} = \frac{1}{2}\left( 1 + V \cos \Delta \phi \right) \ ,
\end{equation}
where the left-hand side is the measured number of counts in port I, normalized by the total counts.\footnote{Note that this is not necessarily the initial number of atoms in an experiment. Atoms that escape the system do not affect the final results, except to reduce the statistical power.}
The main observables are therefore the amplitude $V$, called visibility or contrast, and the phase, $\Delta \phi$. Visibility is a measure of the decoherence of the system, and the phase measures the path differences between the two arms. In an ideal atom interferometer experiment, there would be no decoherence ($V = 1$). However, a loss of visibility, {\it i.e.}~$ V \ne 1$, is common in realistic experiments due to environmental effects and systematics. The visibility $V$ and its quantum noise-limited (QNL) error $\sigma_V^{\rm QNL}$ are obtained by normalizing the expectation value and the variance of the binomial distribution, shown in Eq.~\eqref{eq:exvar}, as
\begin{equation}\label{eqn:sigmaVqnl}
   V = e^{-s}, \quad \text{ and } \quad \sigma_V^{\rm QNL} \equiv \frac{\bar \sigma}{ N_{\rm ind}} = \frac{1}{2\sqrt{N_{\rm ind}}} \; ,
\end{equation}
reflecting the statistics of shot noise fluctuations.

Entangled atom interferometers, matter interferometers, and quantum resonators also exhibit decoherence and phase effects. These experiments are similar to the cold atom interferometers, except that only one independent measurement is made per shot: the quantum state where the single macroscopic object ends. The probability of the object to be measured in port I is also given by Eq.~\eqref{eq:V}, except with $N_{\rm{ind}} = 1$. Note that $s$, in this case, is \textit{not} the decoherence rate per atom, but rather the decoherence rate over the full macroscopic object. In this case, both the real and imaginary part of $\gamma$ should be inferred over many trials~\cite{Riedel2017,Kaltenbaek:2012gt}.

In this paper, we compute the reduction in the fringe amplitude, parametrized by $V$ in Eq.~\eqref{eq:V}, for the cloud (or solid). We note that if the fringe is destroyed by any of these backgrounds, then it is impossible to measure the phase. 

Current matter-wave interferometers have demonstrated detection sensitivity that is limited by the quantum noise limit (see, \textit{e.g.},~\cite{bize2005cold}). For the backgrounds studied here, we will give both the absolute reduction in the amplitude of the fringe, and the corresponding signal-to-noise ratio (SNR) when compared to the QNL, $\sigma_V^{\rm QNL}$ (Eq.~\eqref{eqn:sigmaVqnl}). The SNR determines whether a reduction in amplitude can be detected, given that the QNL quantifies the sensitivity to the amplitude. If we define the amplitude reduction as $\Delta V = V_0 - V$, where $V_0=1$ is the amplitude in the absence of decoherence, then the SNR per shot is:
\begin{equation}
\label{eq:SNRshot}
    \left. \text{SNR}\right|_{\rm shot} \equiv  \frac{|\Delta V|}{\sigma_V } \; ,
\end{equation}
where in this work we take $\sigma_V = \sigma_V^{\rm{QNL}}$. This SNR compares the loss of visibility from a background to shot noise fluctuations. Note that for matter interferometers (\textit{e.g.}, MAQRO) or entangled clouds of atoms, where the target enters the detector as a single object, $N_{\rm ind} = 1$, and $\sigma_V^{\rm QNL} = 0.5$. 

The sensitivity to the signal, and therefore the total SNR, will increase linearly with the exposure time (provided $s \ll 1$) for a background-free experiment. If this holds, $\left. \text{SNR} \right|_{T_{\rm exp}} = \left. \text{SNR} \right|_{\text{shot}} N_{\rm meas}$, where $N_{\rm meas} = T_{\rm exp}/t_{\rm shot}$ is the number of measurements affected per total running time of the experiment, $T_{\rm exp}$. In the presence of a well-characterized background, sensitivity to a small signal can still be achieved over a large background, and in this case, the total SNR scales as $\sqrt{N_{\rm meas}}$.

\begin{figure*}[t]
\centering
\includegraphics[width=0.465\textwidth]{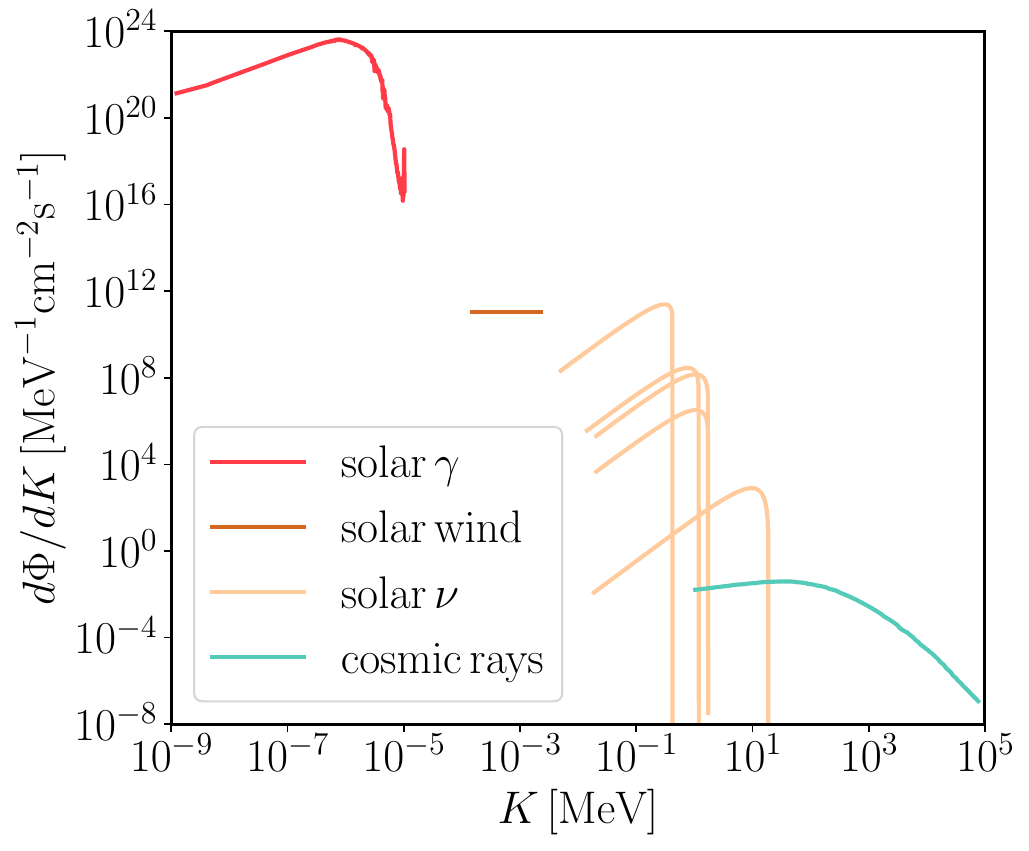}
\includegraphics[width=0.525\textwidth]{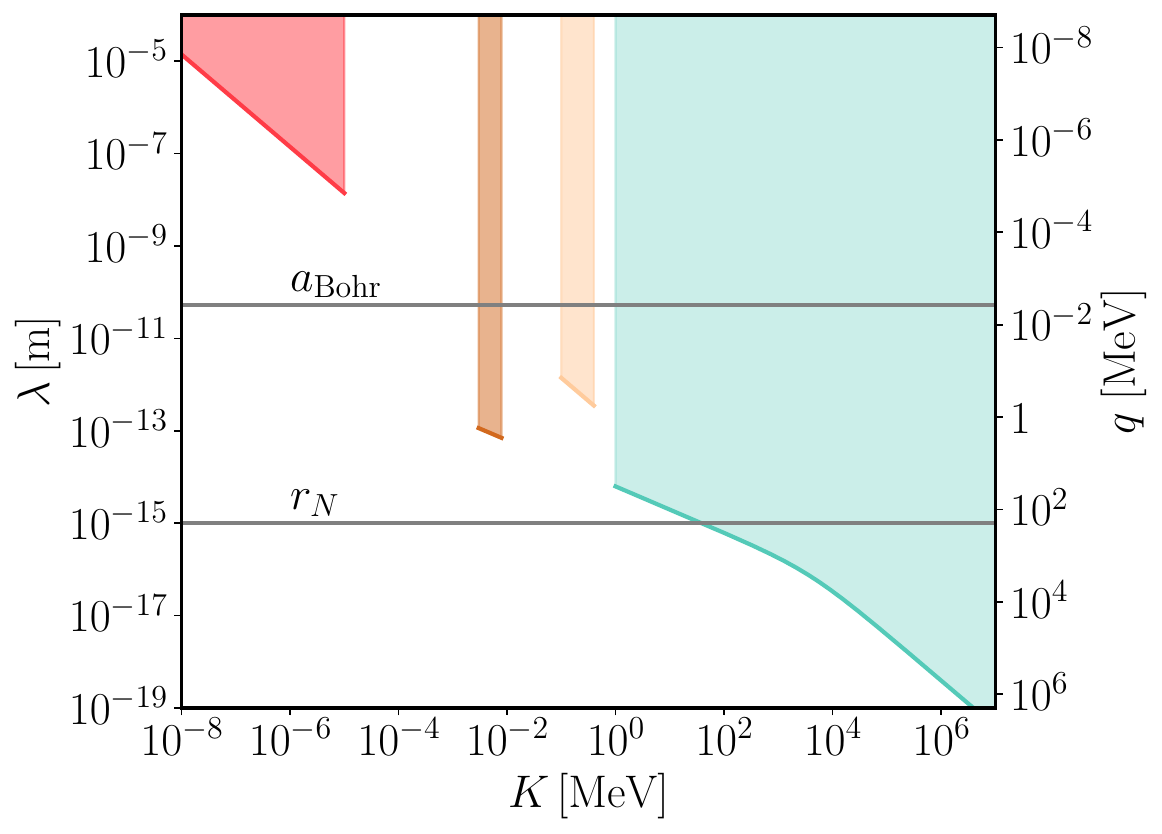}
\caption{Summary of backgrounds relevant for space-based matter-wave interferometer experiments. In the left panel, we show the spectral fluxes ($d\Phi/dK$) and relevant kinetic energies ($K$) of the solar photons (red, \citep{ASTM2000, NREL}), solar wind (brown, \citep{McComas2013}),  solar neutrinos (orange,~\cite{Bahcall:1997eg,RevModPhys.60.297,Bahcall:2004pz}), and galactic cosmic rays (turquoise, \citep{stone2019cosmic}). In the right panel, we show the ``resolution'' ($\lambda = q^{-1}$) of each background to the range of momenta, $q$, that the background can imprint on the target. For elastic scatterings, $q$ can go from zero to its maximum value, given by kinematics (see Eq.~\eqref{eq:qmax}), which determines the shortest wavelength that each background can resolve at a given energy. For charged particles, such as those in the solar wind and cosmic rays, the maximum momentum transferred depends on the mass of the target, which we assume to be protons. The maximum resolution power of the solar photons and neutrinos solely depends on their incoming energy. We do not include the Zodiacal dust due to its inherently different behavior (see Section~\ref{sec:dust}). 
}
\label{fig:kinematics}
\end{figure*}

\subsection{Particle Backgrounds}\label{sec:overview_backrounds}
%%%%%%%%%%%%%%%%%%%%%%%%%%%%%%%%%%
Let us return to the phenomenological picture of atom interferometry, and again consider the decoherence effect. Decoherence from particle scattering in matter-wave interferometers occurs when particles scatter off of one of the wavepackets in an interferometer, and thus ``resolve" the two paths. Let the distance between the two interferometer arms be $\Delta x$, and let the radius of each ``cloud'' (\textit{i.e.,} wavepacket) be $r_{\rm{cloud}}$. The resolving power of an incoming particle is related to its momentum transfer magnitude, $q$. When $q > 1/\Delta x$, the particle is capable of decohering the system (see Fig.~1 of Ref.~\cite{AIDM} for a visual explanation of this relationship). However, if the momentum of the incoming particle is too high, then the particle will begin to resolve individual atoms within the cloud. When these individual atoms are not resolved, $q < 1/r_{\rm{cloud}}$, the rate of interactions may be boosted by the Born enhancement effect -- the rate will scale with the number of targets squared (if they share the same charge)\citep{PhysRevD.98.053004}. The latter will occur when the $s$ in Eq.~\eqref{eqn:sigmaVqnl} corresponds to the decoherence rate over the full macroscopic object. Thus, the most dangerous particle backgrounds for matter interferometers are those that allow for momentum transfers with $1/\Delta x < q < 1/r_{\rm{cloud}}$. %Thus, it is low momentum, high flux particles that could provide the highest backgrounds to our experiments. 
In addition, we note that we only consider elastic scattering processes in this work. Inelastic processes are not coherent, so their rates are very suppressed compared to the coherent elastic processes we consider.

In this paper, we consider a wide variety of possible particle backgrounds for space-based experiments. In Fig.~\ref{fig:kinematics}, we provide an overview of the most salient features for each background. As can be seen from Fig.~\ref{fig:kinematics}, the flux of the background particles is highest for particles with low kinetic energies. Each of our particle backgrounds has vastly different fluxes and interaction mechanisms. The left panel of Fig.~\ref{fig:kinematics} shows the spectral fluxes for the backgrounds we consider, except dust which is treated separately in Section~\ref{sec:dust}. Solar photons dominate the flux. Next, the solar wind and solar neutrinos have similar fluxes, which are many orders of magnitude smaller than the photon flux. Galactic cosmic rays have the smallest flux of the backgrounds we consider.

The energy ranges for each of these backgrounds is set by the processes that produce them and environmental effects. The solar photon spectrum is, roughly, a blackbody with effective temperature set by the temperature of the Sun's photosphere ($T_{\rm{eff}, \odot} = 5772~\rm{K}$ \citep{Prsa2016}). Small deviations from the blackbody are introduced by the chemical composition and physical motions of the Sun's chromosphere. The energy limits are based on the instruments used to measure the solar spectrum, rather than any intrinsic energy limits. The solar wind is composed of charged particles (mostly protons and electrons) that are produced and accelerated by the Sun and its magnetic field. The energy range is set by the velocities and number densities of the particles, but the flux itself is roughly constant \citep{Schwenn1990, MeyerVernet2006, LeChat2012}. Solar neutrinos are byproducts of nuclear fusion in the core of the Sun. As we describe in further detail in Section~\ref{sec:neutrinos},
the maximum energy is set by the mass difference of the other final and initial particles involved in the reactions. For the most relevant reaction chain, $p + p \to d + e + \nu_e$, the endpoint energy of the neutrino is $420.22$ keV, but including the thermal kinetic energy of the protons in the solar plasma raises it to $423.41$ keV~\cite{Bahcall:1997eg}. Solar neutrinos have no minimum energy, but their flux falls according to a power law for low frequencies because of the low energy neutrino phase-space~\cite{Vitagliano:2017odj}. Galactic cosmic rays are charged particles that have been accelerated to high energies by shocks (\textit{e.g.}, from supernovae). The flux of galactic cosmic rays is attenuated by the Sun's solar wind -- this barrier blocks lower energy galactic cosmic rays from entering the solar system. We arbitrarily cut the high end of the galactic cosmic ray spectrum at 100 GeV, which is set by the dataset we use \citep{stone2019cosmic}. However, since the flux only decreases further and these higher energies do not contribute to any scattering rates that are boosted by coherent enhancement, the exact cut-off should not affect our results. Note that this means that we do not consider extragalactic cosmic rays in this paper.

In the right panel of Fig.~\ref{fig:kinematics}, we show the comparison between the range of the elastic scattering $\lambda$ (i.e. the inverse momentum transfer $q^{-1}$), with various characteristic length scales of the matter-wave interferometer systems. The maximum momentum transfer is determined by kinematics, and for elastic scattering it ranges from zero to:
\begin{equation}\label{eq:qmax}
    q_{\rm max} = 2 M \frac{\sqrt{K^2 + 2 m K} (M + m + K)}{(M + m )^2 +  2 M K}~,
\end{equation}
in the laboratory frame. 
Here $K$ is the kinetic energy of the background particle, $m$ is its mass, and $M$ is the mass of the target. While for the atom interferometers the contrast loss does not receive coherent enhancements~\cite{Badurina:2024nge}, for matter interferometers, when the interaction range is longer than the size of the cloud, $r_{\rm cloud}$, individual atoms in the cloud are indistinguishable to a coherent scattering process, and thus the cross-section is Born enhanced by the large number of atoms in the cloud. At the same time, for all interferometers, if the scattering does not resolve the two interfering clouds, which are separated by a distance $\Delta x$, the decoherence is suppressed. This provides a cutoff in the long-range limit of scattering decoherence. For each background, the interaction mechanism depends on the incoming energy of the particles. For example, the scattering range for solar photons is never smaller than the Bohr radius, and thus the solar photon only scatters with the matter-wave interferometer clouds through the polarizability. This is known as Rayleigh scattering for long-wavelength photons. However, because of their heavy mass and relatively high kinetic energy, cosmic ray ion scattering could resolve the atom in some regimes, and thus its interactions would occur through the Coulomb potential.
Solar neutrinos scatter with the cloud through the neutrino-nucleon interaction, and because of their low energies are always in the coherent regime. However, the decoherence effect is very suppressed due to the low solar neutrino flux and low interaction strength. For zodiacal dust, we do not consider specific known interactions, and we instead just take the geometric limit. We provide a detailed discussion of the specific interactions we consider in the following sections.

We note that all of these backgrounds are highly anisotropic for a spacecraft in an Earth-like orbit. However, we assume they are isotropic for the purposes of our calculations in this paper (see appendix in Ref.~\cite{AIDM} to gauge the effect of the daily modulation on the fringe reduction). We revisit and discuss this assumption further in Section~\ref{sec:conclusion}.

\subsection{Space-based Matter-Wave Interferometer Experiments}\label{sec:exps}

In this paper, we consider only space-based matter-wave interferometer mission concepts. We note that all of the experiments are futuristic and do not have finalized experimental parameters. Here we give an overview of each of the experiments and the parameters we assume for each. A summary of this information is given in Table~\ref{tab:exps}.

We note that the flux of background particles will depend heavily on the orbits of the spacecrafts. While some of the experiments have proposed orbits (\textit{e.g.,} BECCAL on the International Space Station \cite{Frye2021}, and AEDGE on a medium Earth orbit \cite{AEDGE}), most of the orbits are unknown. For all of these experiments, we assume an Earth-like orbit. We do not calculate any attenuation or reflection effects from the Earth itself and focus solely on solar and galactic backgrounds that would be present on a spacecraft in a 1 AU orbit about the Sun. In addition, as mentioned above, we do not consider any angular effects -- all backgrounds are taken to be isotropic. We discuss these points further in Section~\ref{sec:conclusion}.

The Bose-Einstein Condensate and Cold Atom Laboratory (BECCAL) is a proposed upgrade \citep{Frye2021} to the Cold Atom Lab (CAL), which is currently running on the International Space Station (ISS) \citep{Elliott2018, Aveline2020}. This upgrade would improve the current atom interferometer capabilities of CAL, including a larger number of atoms in the condensates. Of the experiments we consider, this is the most likely to occur on a short timescale. BECCAL would study a few different atom types; however for simplicity we focus on its $^{87}$Rb capabilities in this paper. We assume that BECCAL would produce condensates with $10^6$ atoms, a cloud size of $150~\mu\rm{m}$, and a free-fall (shot) time of 2.6 s \citep{Frye2021}. The separation of the atom interferometer arms is not given. We assume $\Delta x = 3~\rm{mm}$, which is larger than the cloud sizes and smaller than the total size of the experimental setup. 

The Gravity Probe and Dark energy Detection mission (GDM) is a NASA Innovative Advanced Concepts (NIAC) Phase II mission concept \cite{GDM_NIAC, GDM}. It would consist of a constellation of four spacecraft, each with six atom interferometers. In this paper, we consider the effects on just one of these interferometers. This is a futuristic mission concept and many of the parameters are not concrete yet. We take the same values assumed by the GDM team in their benchmark forecasts \citep{Shengwey}. We assume an atom interferometer baseline of $\Delta x = 25~\rm{m}$, a shot time of $20~\rm{s}$, a cloud size of $r_{\rm{cloud}} = 1~\rm{mm}$, and $10^8$ $^{87}$Rb atoms.

The Atomic Experiment for Dark Matter and Gravity Exploration in Space (AEDGE) is another futuristic mission concept \citep{AEDGE}. It is envisaged as a successor to the long-baseline terrestrial interferometers MAGIS \citep{MAGIS-100:2021etm} and AION \citep{AION}. Like these experiments and GDM, AEDGE would rely on a network of atom interferometers that are separated by very long ($>10^7~\rm{m}$) baselines. As with GDM, we consider just one of AEDGE's atom interferometers in this paper. AEDGE will use $^{87}$Sr atoms that are dropped with a measurement time of $600$~s \citep{AEDGE}. The maximum separation of the atom clouds will be $\Delta x = 90~\rm{cm}$ \citep{AEDGE}. The number of atoms and size of the atomic clouds is not reported in their whitepaper. We assume that their reported phase sensitivity, $10^{-5}~\rm{rad~Hz}^{-1/2}$, is the QNL, and thus scales with the number of atoms via $\delta \phi = 1/\sqrt{N_{\rm{atoms}}}$ \citep{Itano1993, Sorrentino2014}. For AEDGE, the atom clouds would be interleaved and the measurements would occur every $1$~s \citep{Graham2016}. Thus, this trivially converts to the total phase sensitivity per shot, $\delta \phi = 10^{-5}~\rm{rad}$. This then gives us $N_{\rm{atoms}} = 10^{10}$. Likewise, the radius of the cloud is not specified in their whitepaper. We estimate it by assuming that the cloud has a density similar to the atom clouds used in the drop tower experiment at Stanford~\cite{Asenbaum:2020era,Stanford}, which is also a diffuse cold atom interferometer. This then gives a cloud radius of: $r_{\rm cloud}^{\rm AEDGE} = r_{\rm cloud}^{\rm Stanford} (N_{\rm atoms}^{\rm AEDGE} / N_{\rm atoms}^{\rm Stanford})^{1/3} \simeq 3$ mm.

Macroscopic Quantum Resonators (MAQRO) is a proposed space mission to perform high-mass matter interferometry \citep{Kaltenbaek2015, Kaltenbaek2016}. This is the only matter interferometer that we consider in this paper, and thus the only experiment with $N_{\rm{ind}} = 1$. The experiment would use SiO$_2$ molecules \citep{Riedel2017} in a solid sphere with $10^{10}$ nucleons and a radius of $120~\rm{nm}$. The baseline separation would be $100~\rm{nm}$ and the measurement time per shot would be $100~\rm{s}$ \citep{Kaltenbaek2016}.

\begin{table}
    \centering
        \begin{tabular}{|l|c|c|c|c|c|c|}
        \hline
        Experiment & Type of Exp. & Target  & $N_{\rm{nuc}}$  &  $r_{\rm{cloud}}$ [m] & $\Delta x$ [m] & $t_{\rm{shot}}$ [s]\\
       % & & & [m] & & [m] & [s]\\
        \hline
        \hline
        MAQRO \citep{Kaltenbaek2015, Kaltenbaek2016} & Solid & SiO$_2$  & $ 10^{10}$   & $1.2\e{-7}$ & $10^{-7}$ & $100$ \\
        BECCAL \citep{Frye2021, Elliott2018, Aveline2020} & BEC & $^{87}$Rb  & $8.7 \e{7}$   & $1.5\e{-4}$ & $3\e{-3}$ & 2.6 \\
        GDM \citep{GDM, GDM_NIAC, Shengwey}& BEC & $^{87}$Rb  & $8.7 \e{8}$  & $10^{-3}$ & $25$   & $20$\\
        AEDGE \citep{AEDGE} & Diffuse cloud & $^{87}$Sr  & $8.7 \e{11}$  & $3 \times 10^{-3}$ & 0.9 & 600 \\
        \hline
    \end{tabular}
    \caption{Mission concept parameters assumed in this paper. The parameters are taken from the references shown next to each experiment. The details are given in Section~\ref{sec:exps}.
    }
    \label{tab:exps}
\end{table}

\subsection{Summary of Results}
\label{sec:summary}
%%%%%%%%%%%%%%%%%%%%%%%%%%%%%%%

The loss of visibility per shot, $\Delta V$, for each of the backgrounds considered in this work is summarized in Table~\ref{tab:full_res}. The decoherence effects caused by solar neutrinos and zodiacal dust are negligible and is not shown explicitly in the table. Among these backgrounds, the solar wind induces the most significant visibility loss. In particular, the solar wind causes a decoherence background in matter-wave interferometers above the QNL, $\sigma_V^{\rm{QNL}}$, within a single experimental shot. For the matter interferometer mission considered in this work, the solar wind would also destroy the fringe in a single shot and thus preclude any possible measurements. This can be understood due to its relatively high number density (as opposed to galactic cosmic rays), and stronger coupling to the target (as compared to solar photons). Solar photons, with the highest flux, could also destroy the fringe of matter interferometers. For atom interferometers, while the induced contrast loss is below the QNL for each shot, over the full experimental run time, solar photon backgrounds statistically accumulate over the number of total shots (or measurements), $N_{\rm meas}$, and exceed the QNL, thus mimicking a signal without shielding or background rejection. Lastly, the high energy galactic cosmic rays are not a relevant background through elastic scatterings as considered in this work. %This is because of their relatively low flux and high kinetic energy, excluding Born enhancement in most of their kinematic phase space. 
We acknowledge that these high energy particles could have an impact through inelastic processes such as ionization, but this is beyond the scope of this work.

\begin{figure*}[t]
\centering
\includegraphics[width=0.47\textwidth]{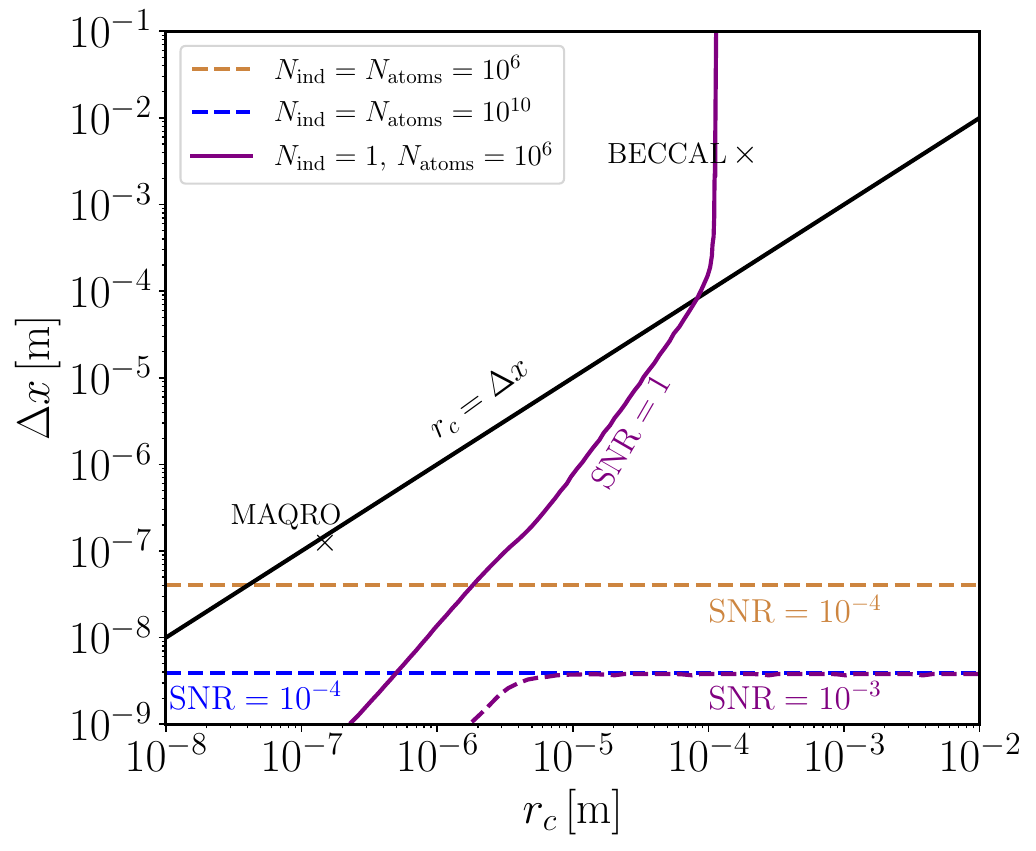}
\includegraphics[width=0.46\textwidth]{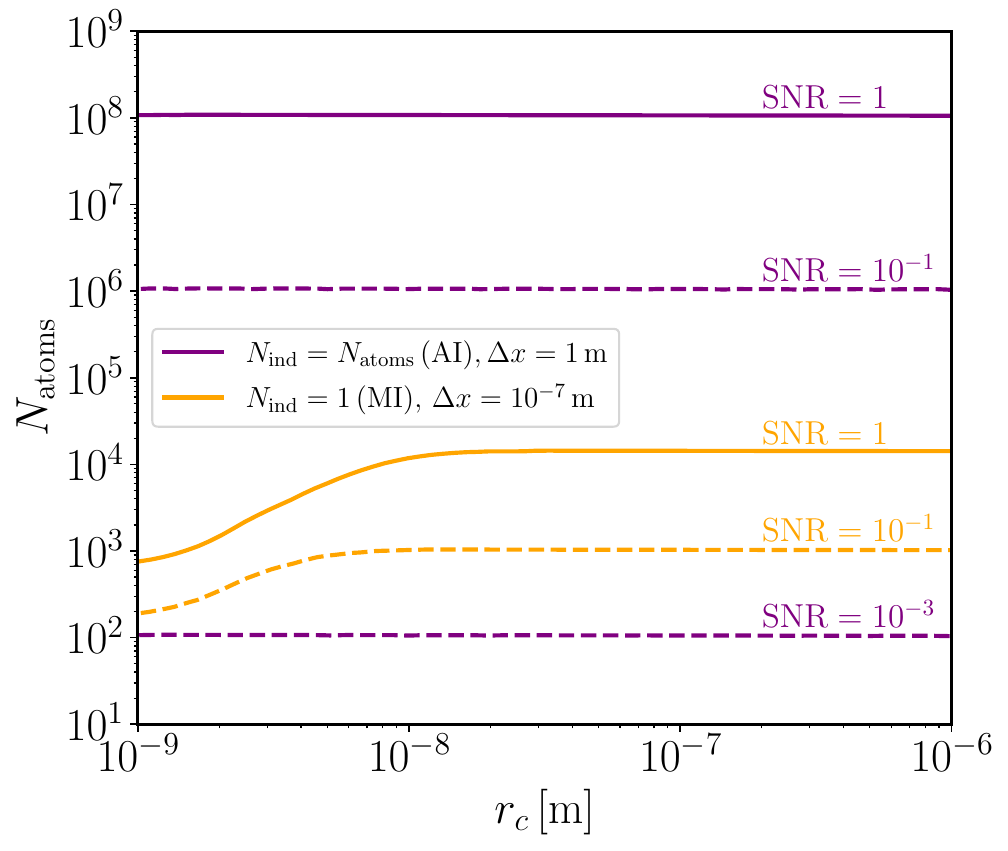}
\caption{The SNR per shot for the solar photon (left) and the solar wind (right) backgrounds as a function of matter-wave interferometer cloud size, $r_{\rm cloud}$, and (left-panel) cloud separation $\Delta x$ for different numbers of atoms $N_{\rm atoms}$, and (right-panel) number of atoms for different $\Delta x$ in matter-wave interferometers. Curves with $N_{\rm ind}=1$ correspond to matter interferometers, while curves with $N_{\rm ind}= N_{\rm atoms}$ assume a cold atom cloud. We have used $t_{\rm shot}=1\,$s and the $^{87}$Rb atom for illustration purposes. Some of the experiments we consider in this paper are shown as benchmark points in the left-panel, marked by `x'; we do not mark GDM and AEDGE whose cloud separation is meter scale and hence not in the plot region. We discuss the limiting behavior governing these curves in Section~\ref{sec:summary}.}
\label{fig:scaling}
\end{figure*}

To aid with the intuition of how these processes affect matter-wave interferometers, we show in~Fig.~\ref{fig:scaling} how the SNR$_{\rm{shot}}$ scales with $\Delta x$ and $r_{\rm{cloud}}$ for the two most relevant backgrounds: the solar wind and solar photons. For illustration purposes, we have chosen $t_{\rm shot}=1$~s, $^{87}$Rb atoms, and different values of $N_{\rm atoms}$ for both atom interferometers ($N_{\rm ind} = N_{\rm atoms}$) and matter interferometers ($N_{\rm ind} = 1$). Atoms are assumed to be uniformly distributed in the target. Note that the $\Delta x < r_{\rm cloud}$ region (below the solid, black line) is generally not implemented experimentally, since the clouds would be overlapping. For missions we consider in this work, only MAQRO has $\Delta x \sim r_{\rm cloud}$. There exist simple scaling laws in some of the regions of the parameter space.

For solar photons with a typical wavelength $\lambda_{\gamma {\odot}}$, in the region where $\Delta x \gg r_{\rm cloud} \gg \lambda_{\gamma {\odot}}$, the differential rate scales roughly as:
\begin{equation}
    d \Gamma_{\rm{tot}, \gamma \odot}/d\omega \propto N_{\rm{atoms}} \lambda_{\gamma {\odot}}^{-4} \, [ N_{\rm{atoms}} \, F_{\rm AI}(r_{\rm cloud}/\lambda_{\gamma {\odot}}) + \mathcal{O}(1) ] \quad \mathrm{for} \:  \Delta x \gg r_{\rm cloud} \gg \lambda_{\gamma {\odot}}\; ,
\end{equation}
where $\omega$ is the photon frequency. In this limit, $F_{\rm AI}( r_{\rm cloud}/\lambda_{\gamma {\odot}}) =  (r_{\rm cloud}/\lambda_{\gamma {\odot}})^{-4}$ for matter interferometers, while for cold atom clouds $F_{\rm AI}(r_{\rm cloud}/\lambda_{\gamma {\odot}}) = 0$. For cold atom clouds, and matter interferometers in the region where $r_{\rm cloud} < \lambda_{\gamma {\odot}}$ or $r_{\rm cloud} \gg N_{\rm atoms}^{1/4} \, \lambda_{\gamma {\odot}}$, the rate only depends on $\Delta x$. Overall, one observes that larger clouds and smaller cloud separations are less affected.

For the solar wind, the rate for Coulomb potential scattering (high momentum transfer) does not depend on either $\Delta x$ or $r_{\rm cloud}$, provided both are greater than the size of an atom. This constant contribution only depends on the type of atom and scales linearly with $N_{\rm atoms}$. On the other hand, the rate for the scattering through polarizability (low momentum transfer), in general, depends on both $\Delta x$ and $r_{\rm cloud}$, the latter when matter interferometers are considered. However, again considering both scales are super-atomic, for $\Delta x > r_{\rm cloud}$, the rate scales as:
\begin{equation}
    \Gamma_{\rm{tot,SW}}^{{\rm low\,}q} \propto N_{\rm{atoms}} r_{\rm atom}^{-4} [ N_{\rm{atoms}} \, F_{\rm AI}(r_{\rm cloud}/r_{\rm atom}) + \mathcal{O}(1)] \quad \mathrm{for} \: \Delta x > r_{\rm cloud} \; ,
\end{equation}
 which describes the behavior of the contours on the right panel of~Fig.~\ref{fig:scaling}. In this limit, $F_{\rm AI}(r_{\rm cloud}/r_{\rm atom})=(r_{\rm cloud}/r_{\rm atom})^{-4}$ for matter interferometers, while $F_{\rm AI}(r_{\rm cloud}/r_{\rm atom})=0$ for cold atomic clouds. Notice that, for matter interferometers, the rate is independent of $r_{\rm cloud}$ and $\Delta x$ when $r_{\rm cloud} \gg N_{\rm atoms}^{1/4} \, r_{\rm atom}$.  

The overall picture is that adequate shielding of both the solar wind and solar photons will be important for the success of future space-based matter-wave interferometers. We now turn to describe our methods and results in detail.

\begin{table}
\begin{tabular}{ |c ||c | c | c || c|| c | }
\hline	
Exp./Bkgs. & Solar Photons & Solar Wind & Galactic Cosmic Rays &  $\sigma_V^{\rm QNL}$ & $N_{\rm meas}$ \\
\hline
\hline
MAQRO &  $1$ \qquad \qquad \qquad \  [$2$] &  $1$ \qquad \qquad \quad \ \ \ [$2$]& $3 \times 10^{-7}$ \ \  [$6 \times 10^{-7}$] & $0.5$ & $3 \times 10^5$  \\
BECCAL &  $1 \times 10^{-6}$ [$2 \times 10^{-3}$]  &   $2 \times 10^{-4}$ \quad \quad [$0.4$] & $1 \times 10^{-16}$ [$2 \times 10^{-13}$]  &  $5 \times 10^{-4}$ & $ 10^7$   \\
GDM &  $1 \times 10^{-5}$ \qquad \ [$0.2$]  &   $1 \times 10^{-3}$ [$2 \times 10^1$]  &  $7 \times 10^{-16}$ [$1 \times 10^{-11}$] & $5 \times 10^{-5}$ & $2 \times 10^6$   \\
AEDGE & $9 \times 10^{-5}$  \ \ [$2 \times 10^{1}$]  & $3 \times 10^{-2}$ [$6 \times 10^{3}$] & $2 \times 10^{-14}$ \ [$4 \times 10^{-9}$] & $5 \times 10^{-6}$ & $5 \times 10^4$  \\
\hline
\end{tabular}
\caption{Loss of visibility $\Delta V \equiv 1 - V$ over a single experimental interrogation time $t_{\rm shot}$ as caused by scatterings with various particle backgrounds in the space environment. The decoherence caused by solar neutrinos and Zodiacal dust is negligible. The numbers are obtained assuming no shielding. For comparison with the experimental sensitivity, we list in squared brackets the SNR$_{\rm shot}$ defined in Eq.~\eqref{eq:SNRshot}, where we assume the noise is dominated by quantum shot noise $\sigma_V^{\rm QNL}$ (second to last column, Eq.~\eqref{eqn:sigmaVqnl}).  
Depending on the measurement scheme, different statistics on the number of measurements $N_{\rm meas} = T_{\rm exp} / t_{\rm shot}$ (last column) over the experiment run time should be further applied to $\left. \text{SNR}\right|_{\rm shot}$, as discussed below Eq.~\eqref{eq:SNRshot}. We have assumed $T_{\rm exp}=1$ year and an isotropic flux consistent with a 1 AU orbit for all missions. The details of each experiment are given in Table~\ref{tab:exps} and Section~\ref{sec:exps}. 
} 
\label{tab:full_res}
\end{table}

%%%%%%%%%%%%%%%%%%%%%%%%%%%%%%%%%%%%%%%%%%%%%%%%
\section{General Formalism}\label{sec:formalism}
%%%%%%%%%%%%%%%%%%%%%%%%%%%%%%%%%%%%%%%%%%%%%%%%
In this section, we summarize the formalism that we use to calculate the decoherence induced by the particle backgrounds we consider. We largely follow the formalism outlined in Ref.~\cite{AIDM}, which builds on the formalism in Refs.~\cite{Coskuner:2018are,Trickle:2019nya}. However, unlike in the dark matter case, we do not assume a constant density of particles at all energies. In addition, we must modify the mediator and velocity phase space terms for the particular particles and interactions we consider. 

We begin with the rate of decoherence-inducing interactions. This is a function of the number density and energies of the incoming particles, their scattering interactions with the cloud\footnote{In the case of atom interferometers, the relevant object for the scattering rate is the cloud of atoms. For matter interferometers or resonators, the relevant object is the solid. In the following, we refer to both as ``clouds''.}, and the response of the cloud itself. We can write the differential decoherence rate 
for the entire cloud
as:
\begin{equation}
\begin{split}
\frac{d\, \Gamma_{\rm tot}}{d\omega}
= 
V
\int d\Omega \frac{dn(\omega, \Omega)}{d\omega \, d\Omega}  \Gamma
(\omega,\Omega),
\end{split}
\end{equation}
where $n$ is the number density of the incoming particle, and $\omega$ and $\Omega$ are the energy and solid angle of the incoming particle, respectively.
All of the physics of the interactions lies in the decoherence rate per interaction, 
$\Gamma (\omega,\Omega)$.
Applying Fermi's Golden rule, this reads: 
\begin{equation}\label{eq:GammaPre}
    \begin{split}
\Gamma
(\omega,\Omega) 
&= \frac{1}{V} \int \frac{d^3 {\bf q} }{(2 \pi)^3} \overline{|\sf{M}|^2} \, p_{\rm{dec}}({\bf q \cdot \Delta x}) \, S(\mathbf{q}) \, 2 \pi \delta (E_f - E_i - \omega_{\bf q}) \; , \\
    \end{split}
\end{equation}
where the momentum transfer is ${\bf q} = {\bf p'} - {\bf p}$, and the energy deposition is $\omega_{\bf q} = \omega' - \omega$. $\omega'$ denotes the energy of the outgoing particle (see the diagram below for the definition of the quantities relevant for the kinematics of the scattering process). 
\begin{equation}
\begin{gathered}
\begin{tikzpicture}[line width=1.5 pt,node distance=1 cm and 1 cm]
\coordinate[label=left:${\rm background} \, \{ m{\rm ,} \, (\omega {\rm ,} \, \mathbf{p}) \}$] (i1);
\coordinate[right = 1 cm of i1](aux1);
\coordinate[below = 0.5 cm of aux1](v1);
\coordinate[right = 1 cm of aux1,label=right:${\rm background} \, \{ m{\rm ,}\, ( \omega' {\rm ,} \, \mathbf{p'})\}$] (i2);
\coordinate[below = 0.5 cm of v1](aux2);
\coordinate[left = 1 cm of aux2,label=left:${\rm target} \, \{ M {\rm ,} \,( E_i {\rm ,} \, \mathbf{k}) \}$] (i3);
\coordinate[right = 1 cm of aux2,label=right:${\rm target} \, \{ M {\rm ,}\,( E_f {\rm ,} \, \mathbf{k'})\}$] (i4);

\draw[fermion] (i1)--(v1);
\draw[fermion](v1)--(i2);
\draw[fermion](i3)--(v1);
\draw[fermion](v1)--(i4);

\draw[fill=white](v1) circle (.405 cm);
\draw[pattern=north east lines](v1) circle (.4 cm );

\end{tikzpicture}
\end{gathered}
\end{equation}
$\overline{|\sf{M}|^2}$ is the spin-averaged scattering matrix element squared, with the quantum states unit normalized, {\it e.g.} $\langle \mathbf{p} | \mathbf{p} \rangle = \langle i | i \rangle = 1$~\cite{Trickle:2019nya}.\footnote{We use the matrix element for unit normalized quantum states, $\sf{M}$, to follow the conventions used in Ref.~\cite{AIDM}. It relates to the matrix element, ${\cal M}$, in the standard normalization ($\langle p_k | p_k \rangle = 2 E_k$) by $\overline{|\sf{M}|^2} = \overline{|{\cal M}|^2} / (2E_i \, 2 E_f \, 2 \omega \, 2 \omega')$. See~Appendix~\ref{app:formalism} for more details.} The probability that a background produces decoherence in a single atom is given by~\cite{Riedel:2012ur,PhysRevD.96.023007,Hornberger_2003,Joos:1984uk,AIDM}: 
\begin{equation}\label{eq:pdec}
    p_{\rm dec} = {\cal R}{\rm e}\{1-  {\rm exp}[i \mathbf{q} \cdot \mathbf{\Delta x}]\} \; .
\end{equation}
The cloud response to the scattering, depending on which type of measurement is performed, is encoded in the static structure function $S(\mathbf{q})$. The latter sums over the constituents of the cloud, which we refer to as ``targets'' of the interaction ({\it e.g.} atoms, nucleons, electrons inside the cloud, or even the cloud itself, depending on the interaction), that scatter with the background particles. The potential Born enhancement effects are contained within this term: $S(\mathbf{q})$ will scale as $N_{\rm target}$ for contrast loss in cold atom clouds when 1-body measurements are performed, while matter interferometers will get its contrast loss enhanced by $N_{\rm target}^2$~\cite{Badurina:2024nge}. Note that since the directional information of the target is included in $p_{\rm dec} ({\bf q \cdot \Delta x})$, the structure function only depends on the magnitude of the momentum transfer.

Throughout this paper, we adopt the laboratory frame and choose the $\hat{\mathbf{z}}$ direction to be aligned with the momentum of the incoming particle $\mathbf{p}$. In such a frame, the target is initially at rest ($E_i = M$, $\mathbf{k}=\mathbf{0}$) and is kicked by some background particle, which transfers a momentum $\mathbf{q}$ to the target, making it recoil with a final energy $E_f = \sqrt{q^2 + M^2}$. Notice that the recoil energy $E_f - M$ is suppressed, since $q \ll M$. Thus, we assume that the targets are non-recoiling in this formalism.

According to momentum and energy conservation, the magnitude of the momentum transfer, $q \equiv \| {\bf q} \|$, can be determined by the background particle's mass, $m$, the target particle's mass $M$, the kinetic energy $K$, and the angle between $\mathbf{p}$ and $\mathbf{q}$. In the lab frame, $0 \le q \le q_{\rm max}$, where $q_{\rm max}$ is defined in Eq.~\eqref{eq:qmax}.
As discussed in Sec.~\ref{sec:overview_backrounds}, $q_{\rm max}$ determines the shortest scattering scale allowed by the kinematics.

In general, the flux of incoming particles is time-dependent, due to the motion of the spacecraft. 
For simplicity, in this work, we omit this time dependence, and thus the daily/orbit modulation of the decoherence. 
As we discuss in Section~\ref{sec:overview_backrounds}, we further assume that the background particles are isotropic. 
Thus, the angular part of the phase space can be integrated out. We refer the reader to Appendix~\ref{app:ang} for the explicit integration of the angular part.
With these assumptions, the decoherence per measured object,
$s$, over a measurement time per shot, $t_{\rm shot}$, in the lab frame is:
\begin{equation}\label{eq:master}
    \begin{split}
s &= t_{\rm shot} \, \frac{\Gamma_{\rm tot} }{N_{\rm ind}}\\
&= \frac{t_{\rm shot}}{2 \pi \, N_{\rm ind}}   \int d\omega \frac{dn(\omega)}{d\omega} \int dq \, q\frac{M + \omega- \sqrt{M^2+q^2}}{\sqrt{\omega^2 - m^2}} \, \overline{|{\sf M} (q,\omega)|^2} \,  S(q) \, \Big( 1 - {\rm sinc} (q \Delta x) \Big) \; ,
\end{split}
\end{equation}
where $N_{\rm ind}$ is the number of independent objects that are measured. For a solid, $N_{\rm ind} = 1$, and for a cold atom cloud, $N_{\rm ind} = N_{\rm atoms}$.
%Notice that, in the mean-field approximation, the right-hand side of \redit{Eq.~\eqref{eq:GammaPre}} is always real, implying that $\phi_{\rm iso} = 0$ for isotropic sources (see Ref.~\cite{AIDM} for a longer discussion of this point). 
Note that the spectral density can also be written in terms of the flux (number of particles per area per time). In Appendix~\ref{app:formalism}, we give the derivation of the general rate formula in terms of the cross section and the flux of incoming particles.

\section{Solar Photons}\label{sec:photons}
%%%%%%%%%%%%%%%

As the right panel of Fig.~\ref{fig:kinematics} shows, solar photons have the highest flux of the backgrounds that we consider, and their energies are not high enough to resolve the atoms in the clouds. In this section, we compute the decoherence associated with solar photon scattering in matter-wave interferometers. We first discuss the solar flux data, and we then derive the relevant matrix element for photon scattering with atoms.

For our flux calculations, we use the solar irradiance (flux/surface area) data from Ref.~\citep{ASTM2000}, accessed via Ref.~\citep{NREL}. This spectrum gives the spectral irradiance of solar photons at 1~AU for zero air mass ({\it e.g.}, outside of Earth's atmosphere). It is calculated by combining measurements from various space-based, balloon-borne, sounding rocket, and ground-based experiments. The integrated solar irradiance is normalized to the solar constant value of $1366.1~\rm{W/m}^2$. Note that there are other solar spectra that have been compiled, but they are mostly within 5\% of this spectrum~(see, {\it e.g.}, \cite{Gueymard2018}). The data is given as a spectral irradiance in $\rm{W/(m}^2\mu\rm{m})$, which we convert to $\rm{MeV}^{-1}\rm{cm}^{-2}\rm{s}^{-1}$. The resulting flux is shown in the left panel of Fig.~\ref{fig:kinematics}.

Although atoms have neutral electromagnetic charge, they can acquire an electric dipole moment proportional to an applied electromagnetic field. The tendency of atoms to acquire a dipole moment is parameterized by their polarizability. The scattering process of photons with atoms through this polarizability is known as Rayleigh scattering. Therefore, we compute the interaction of a solar photon with the cloud atoms through the following Hamiltonian:
\begin{equation}
H_I = - \frac{1}{2} \int d^3 \mathbf{r} \, \mathbf{P}(\mathbf{r}) \cdot \mathbf{E}_{\rm out}(\mathbf{r}) =  -\frac{1}{2}\alpha_N \int d^3 \mathbf{r}  \, n(\mathbf{r}) \, \mathbf{E}_{\rm in}(\mathbf{r}) \cdot \mathbf{E}_{\rm out}(\mathbf{r}) \; ,
\label{eq:HI}
\end{equation}
where $\mathbf{E}_{\rm in}(\mathbf{r})$ ($\mathbf{E}_{\rm out}(\mathbf{r})$) represents the incoming (outgoing) solar photon. We assume linear polarizability induced solely by the electric field of the incoming photon, thus $\mathbf{P}(\mathbf{r}) = \alpha_N  \mathbf{E}_{\rm in}(\mathbf{r})$, where $\alpha_N  = 4 \pi  r_{\rm atom}^3 \epsilon_0 (\epsilon_r -1)/(\epsilon_r +2 )$. Here, $r_{\rm atom}$ is the size of the scattered atom, $\epsilon_0$ is the vacuum permittivity, and $\epsilon_r$ is the dielectric of the target. The electric fields are quantized as follows:
\begin{eqnarray}
\mathbf{E}_{\rm in}(\mathbf{r}) &=& \frac{i}{\sqrt{2V}} \sum_{\mathbf{p}} \sum_\lambda \sqrt{\omega} \left(\mathbf{\epsilon}(\mathbf{p},\lambda) a_{\mathbf{p},\lambda} e^{i \mathbf{p} \cdot \mathbf{r}} - \mathbf{\epsilon}^*(\mathbf{p},\lambda)a^\dagger_{\mathbf{p},\lambda}e^{-i\mathbf{p}\cdot \mathbf{r}}\right),\\
\mathbf{E}_{\rm out}(\mathbf{r}) &=& \frac{i}{\sqrt{2V}} \sum_{\mathbf{p'}} \sum_\lambda \sqrt{\omega'} \left(\mathbf{\epsilon}(\mathbf{p'},\lambda) a_{\mathbf{p'},\lambda} e^{i \mathbf{p'} \cdot \mathbf{r}} - \mathbf{\epsilon}^*(\mathbf{p'},\lambda)a^\dagger_{\mathbf{p'},\lambda}e^{-i\mathbf{p'}\cdot \mathbf{r}}\right) \; .
\end{eqnarray}
The target density can be quantized as:
\begin{equation}
n(\mathbf{r}) = n_0 + V^{-1/2} \sum_\mathbf{q} n_{\mathbf{q}} e^{i \mathbf{q} \cdot \mathbf{r}} \; .
\end{equation}

Applying the quantized electric fields and density to Eq.~\eqref{eq:HI}, 
\begin{equation}
    \begin{split}
 \sum_{f,\mathbf{p}'} |\langle \mathbf{p}', f | H_{I} | \mathbf{p} ,i \rangle|^2 
 &= \frac{1}{V} \int \frac{d^3 \mathbf{q}}{(2 \pi)^3} \frac{1}{16} \alpha_N^2 \omega \, \omega' 
S(q) ,
\end{split}
\end{equation}
we obtain the matrix element $\overline{|\sf{M}|^2} = \alpha_N^2 \omega \omega' /16$. Finally, the total decoherence rate is:
\begin{equation}\label{eq:ratesolarphotons}
    \begin{split}
\left. \Gamma_{\rm tot} \right|_{\gamma_{\odot}}
& = \frac{\alpha_N^2}{32 \pi}  \int d\omega \frac{d n(\omega)}{d \omega} \,\omega \int dq  \Big(\omega + M - \sqrt{M^2 + q^2} \Big) \, q \,   S(q) \,
 \Big[ 1 - {\rm sinc} ( q \Delta x ) \Big] \; ,
    \end{split}
\end{equation}
where the structure-function parametrizes the response of the $N_{\rm atoms}$ atoms in the cloud:
\begin{equation}
    \begin{split} \label{eq:cloud-SF}
S(q) = N_{\rm atoms} + N_{\rm atoms}^2 F_{\rm AI}( q r_{\rm cloud})^2 \; .
\end{split}
\end{equation}
$F_{\rm AI}$ will depend on the nature of the interferometer and the type of measurement. If the atom cloud is initially prepared as a Bose-Einstein Condensate (BEC), such as BECCAL or GDM, or a diffuse cloud, such as AEDGE, no coherent enhancement is expected when 1-body measurements (atom counts in the respective ports) are performed~\citep{Badurina:2024nge}, so that $F_{\rm AI}(q r_{\rm cloud}) = 0$.
For matter interferometers or entangled clouds, however, Born enhancements will occur when $q \ll 1/r_{\rm cloud}$. We assume the atoms are uniformly distributed, and thus use the form factor \cite{Coskuner:2018are}:
\begin{equation} \label{eq:HelmetFF}
    F_{\rm AI}(q r_{\rm cloud}) = \left\{
                \begin{array}{ll}
                  0 \qquad \quad \qquad \ \, \text{for BECCAL, GDM, AEDGE,}\\
                 \displaystyle  \frac{3j_1(q r_{\rm cloud})}{q r_{\rm cloud}} \quad \text{for MAQRO}\; ,
                \end{array}
              \right. 
\end{equation}
where $j_1$ is the spherical Bessel function of the first kind. The polarizability values for the relevant materials are: $\alpha_{\rm Rb} =  47.39\, (8) \, $\AA$^3$ \cite{gregoire2015measurements}, $\alpha_{{\rm SiO}_2} =4.84 \, $\AA$^3$ \cite{lasaga1982electronic}, and $\alpha_{\rm Sr} =  192.40 \, a_0^{3}$ \cite{PhysRevA.89.022506}, where $a_0$ is the Bohr radius.

\begin{table}
\begin{tabular}{|c || c  c c  c  | }
\hline
$\gamma_{\odot}$ & MAQRO  & BECCAL  & GDM    & AEDGE \\
\hline
\hline
Rate [$\text{s}^{-1}$] & $ \ \ 1.2 \times 10^{7} \ \ $ & $\ \ 0.41 \ \ $ & $ \ \ 4.1 \times 10^1 \ \ $ & $1.5 \times 10^3$ \\
\hline
\end{tabular}
\caption{Total decoherence rate $\Gamma_{\rm tot}$ caused by solar photons, in $s^{-1}$.}
\label{tab:solarphotons}
\end{table}

In Table~\ref{tab:solarphotons}, we show the decoherence rate caused by solar photons for the different space missions considered in this work. The visibility for most of the experiments receives a reduction similar to the QNL error from the action of solar photons. In particular, MAQRO is severely affected compared to the other proposals. This enhancement is expected due to the smaller cloud radius and the larger number of atoms contained in MAQRO, the latter contributing to Born enhancements. %\yd{Comment: since the solar photon energy range shown in Fig.1 doesn't resolve the cloud for most of the experiments other than MAQRO, I think we can keep the table, but perhaps delete the rest of this paragraph.} This can be understood by having a close look at the rate in Eq.~\eqref{eq:ratesolarphotons}. The $q$ integral is dominated by the Born $N_{\rm atoms}^2$ enhancement since for most of the missions $\Delta x \gg r_{\rm cloud}$ (for MAQRO $\Delta x \sim r_{\rm cloud}$, but the same reasoning applies). Due to the $q$ scaling, the largest contribution occurs at $q \sim 1/r_{\rm cloud}$. Hence, one can estimate the relative strength of the rate between different experiments by taking the ratio of the scaling of the rate with the parameters of the experiment, $ds/dt \propto \alpha^2 N_{\rm atoms}^2 / r_{\rm cloud}^2$, which roughly justifies the results shown in Table~\ref{tab:solarphotons}.

\section{Cosmic Rays and the Solar Wind}\label{sec:crs}
%%%%%%%%%%%%%%%%%%%%%%%%%%%%%%%%%%
Charged particle backgrounds in space at energies lower than ${\sim}1~\rm{GeV}$ are mainly sourced by the Sun and galactic processes. Charged particles from the Sun, {\it i.e.}~the solar wind, have a narrow energy spectrum with low kinetic energy, $0.3 - 3\,$keV. The spectral density of the solar wind is approximately constant in this kinetic energy window; however, it can have large time-dependent fluctuations because of solar activity. Galactic cosmic ray particles, on the other hand, have relatively high kinetic energies, in the MeV - TeV range. 

Galactic cosmic rays are composed of $\gtrsim 90\%$ protons, $\lesssim 10\%$ alpha particles, and $\ll 1\%$ other particles. The solar wind has about an equal number of electrons as protons, where they have approximately the same velocity \citep{Heiken1991}. Galactic cosmic rays can also include electrons, and their spectral flux is higher than protons at low kinetic energies \citep{stone2019cosmic}. However, we get similar or weaker decoherence rates for electrons than for the protons. Therefore, we only explicitly calculate the proton background, which should approximately capture the impact of these backgrounds. 
For a given kinetic energy and scattering angle, there are two separate interaction types that can dominate. These cases will depend on whether the proton resolves the atom (momentum transfer $q > 1/r_{\rm atom} \sim {\cal O}(\text{MeV})$) or not. We first discuss the flux data we use for both types of particles. We then discuss the interactions that dominate these two momentum transfer limits. Finally, we discuss how we combine these limits to calculate a total rate.

The solar wind has a roughly constant energy flux \citep{Schwenn1990, MeyerVernet2006, LeChat2012}. We use the solar wind parameters for the most recent epoch of 2009--2013, provided in Ref.~\citep{McComas2013}. This gives proton velocities of $v=398~\rm{km/s}$, and proton number densities of $n = 5.74~\rm{cm}^{-3}$. Assuming the solar wind being all protons at low kinetic energies (keV), we convert this to an irradiance of: $343~\rm{W/m}^2$, which is within an order of magnitude of the commonly accepted value, $70~\rm{W/m}^2$. We assume that the flux is constant over the range 0.3--3 keV~\citep{Heiken1991}. We note that these numbers roughly agree with the values that can be found using real-time data from the DSCOVR satellite~\citep{DSCOVR} via Ref.~\citep{RTSW}. In addition, we ignore any coronal mass ejection events in setting these numbers. These would produce much higher energy particles, but these are short events that could be excluded from matter-wave interferometer data collection.

For the flux of galactic cosmic rays, we use theoretical models fit to the Voyager $2$ data \cite{stone2019cosmic}, as plotted in the right-handed panel of Fig.~\ref{fig:kinematics}. We note that this data is taken from beyond the heliopause ({\it e.g.}, outside of the protective influence of the Sun's magnetic field). Thus, the flux is slightly higher than the flux found within the solar system (see {\it e.g.}, Ref.~\cite{Cummings2016} for a discussion of this point). However, this is the highest quality data available at these energy levels that is outside of the Earth's magnetic field, which would also distort the measured galactic cosmic rays. Depending on the orbit of each mission, the galactic cosmic ray flux would vary according to these considerations. 

\subsection{Charged Particle Scattering with Low $q$}

%%%%%%%%%%%%%%%%%%%%%%%

When $q \ll 1/ r_{\rm atom}$, the scattering of charged particle backgrounds does not resolve individual atoms. Therefore, similarly to the case of solar photons, charged particles interact with the neutral atoms in the matter-wave interferometers through their polarizability $\alpha_N$. For protons, since the flux drops exponentially above $1\,$GeV, we assume that the incoming and outgoing charged particles are non-relativistic, and so can then be described by classical waves according to the Born approximation:
\begin{equation}
    \begin{split}
\psi_{i,f} = \frac{1}{\sqrt{V}} e^{i \mathbf{p} \cdot \mathbf{r}}, \frac{1}{\sqrt{V}} e^{i \mathbf{p}' \cdot \mathbf{r}} \; .
    \end{split}
\end{equation}
Atoms in the cloud will be polarized by the Coulomb field of the charged particles, which has an interaction potential of:
\begin{equation}
    \begin{split}
U( \mathbf{r}) = 
- \sum_{l = 1,\cdots, N_{\rm atoms}} \frac{\alpha_N \, Z_{\rm{ion}}^2\, e^2}{ 2 (4 \pi)^2\, \| \mathbf{r} - \mathbf{r}_l \|^4} 
\; , \\
    \end{split}
\end{equation}
where $Z_{\rm{ion}}$ is the charge number of the charged particle, and $\mathbf{r}_l $ is the location of the atoms. According to Fermi's golden rule:
\begin{eqnarray}
  \sum_{f,\mathbf{p}'} |\langle f, \psi_f | H_{I} | i ,\psi_i  \rangle|^2 
 &=& \frac{1}{V}\frac{ \pi^2 Z_{\rm{ion}}^4 \alpha_N^2 \alpha^2 }{64} \int \frac{d^3 q}{(2 \pi)^3}
q^2 S(q)  \; ,
\end{eqnarray}
where $\alpha = e^2 / 4\pi$ is the fine-structure constant, we obtain the matrix element $\overline{|\sf{M}|^2}  = \pi^2 Z_{\rm{ion}}^4 \alpha_N^2 \alpha^2 q^2 /64$. The structure-function $S(q)$ reflects the sum of all atoms in the cloud and would take the same form as given by Eqs.~\eqref{eq:cloud-SF} \& \eqref{eq:HelmetFF}, depending on the experiment type.
Using Eq.~(\ref{eq:master}), the total decoherence rate is then:
\begin{equation}
\begin{split}
   \left. \Gamma_{\rm tot} \right|_{\textsc{cpb},{\rm low}\, q}
   &= \frac{\pi}{128} Z_{\rm{ion}}^4 \alpha^2 \alpha_N^2 \int d\omega \frac{dn}{d\omega} \frac{1}{\sqrt{\omega^2 - m^2}} \\
   & \quad \times \int dq \, \Big(\omega + M -  \sqrt{M^2 + q^2} \Big) q^3 \, S (q) \,\Big[ 1 - {\rm sinc} ( q \Delta x ) \Big] \; ,
   \end{split}
\end{equation}
where the subscript $\textsc{cpb}$ refers to charged particle backgrounds.

\subsection{Charged Particle Scattering with High $q$}
%%%%%%%%%%%%%%%%%%%%%%%%%%%%%%%%%%%%

We now describe the interactions of cosmic ray charged particles that resolve individual atoms ($q \gg 1/r_{\rm atom}$) within the matter-wave interferometer clouds. The matrix element squared of this process is given by: 
\begin{equation}\label{eq:M2CR}
\begin{gathered}
\begin{tikzpicture}[line width=1.5 pt,node distance=1 cm and 1 cm]
\coordinate[label=left:$p^+$] (i1);
\coordinate[below right = 1 cm of i1](aux1);
\coordinate[above = 0.3 cm of aux1](v1);
\coordinate[above right = 1 cm of aux1,label=right:$p^+$](i2);
\coordinate[below = 0.75 cm of v1](v2);
\coordinate[below = 0.45 cm of v1](aux2);
\draw[fermion] (i1)--(v1);
\draw[fermion](v1)--(i2);
\draw[fill=black] (v1) circle (.05cm);
\draw[fill=black] (v2) circle (.05cm);
\coordinate[below left = 1 cm of aux2,label=left:$p^+/e^-$](i3);
\coordinate[below right = 1 cm of aux2,label=right:$p^+/e^-$](i4);
\draw[fermion](i3)--(v2);
\draw[fermion](v2)--(i4);
\draw[vector](v1)--(v2);
\end{tikzpicture}
\end{gathered} \quad 
\begin{gathered}
 \quad \overline{|{\cal M}|^2} = 2(4\pi)^2\alpha^2 Z_{\rm{ion}}^2 
 \frac{2(m^2 + M^2-s)^2
 + 2 st +t^2}{t^2} \; .
\end{gathered}
\end{equation}
In this regime of $q$, the target can be either an electron or a proton. In the following we focus on protons as targets. Similar results are obtained when considering the electrons instead. 

In the laboratory frame, we can write the Mandelstam variables as functions of the masses and the momentum transfer:
\begin{equation}
    t = 2 M^2 (1-\sqrt{1+q^2/M^2}) \quad \text{ and } \quad s=m^2+M^2 + 2 \omega M \; .
\end{equation}
From Fig.~\ref{fig:kinematics}, we can see that the highest flux is dominated by non-relativistic cosmic rays. Thus, assuming protons are the target, $t \sim -q^2$. Expanding in small $q$ ($q \ll m,M$), the matrix element simplifies to: 
\begin{equation}\label{eq:nonrelM2}
    \overline{|{\cal M}|^2} \simeq \frac{4^4 \pi^2 \alpha^2 Z_{\rm{ion}}^2 \omega^2 M^2}{q^4} \; , 
\end{equation}
which matches the scattering amplitude that would be obtained from the cosmic rays interacting with a Coulomb potential $U(r) = e/(4\pi r)$, whose Fourier transform is $\tilde U(q) = e/q^2$.
Converting Eq.~\eqref{eq:nonrelM2} to $|{\sf M}|^2$ by unity normalizing the states (see Eq.~\eqref{eq:conversion} in Appendix~\ref{app:formalism}) and applying it to Eq.~\eqref{eq:master}, we obtain the total decoherence rate caused by cosmic rays interacting with protons when $q \gg 1/r_{\rm atom}$: 
\begin{equation}\label{eq:rateCR}
 \begin{split}
\left. \Gamma_{\rm tot} \right|_{\textsc{cpb},{\rm high} \, q}^{} 
& \simeq
8 \pi Z_{\rm ion}^2 \alpha^2 \int d\omega \frac{d \Phi (\omega)}{d\omega} \frac{\omega^2}{\omega^2 - m^2} \int_{0}^{q_{\rm max}(\omega)} dq \,  \frac{S(q)}{q^3}  
 \Big[ 1 - {\rm sinc} ( q \Delta x ) \Big] \; ,
\end{split}
\end{equation}
where $d\Phi / d \omega$ is the spectral flux.
The structure-function reflects the sum of all protons in the target, and thus takes the form:
\begin{equation}
    \begin{split}
S (q)  = N_{\rm atoms} Z_{\rm{target}} \Big( 1 + Z_{\rm{target}} {F}^2_{\rm N} (q \, r_N)\Big) \; ,
    \end{split}
\end{equation}
where $Z_{\rm{target}}$ is the atomic number of the target, and $F_N(q \, r_N)$ is the nucleus form factor, which is given by the Helm form factor~\cite{Helm:1956zz}:
\begin{equation} \label{eq:FN}
    F_{\rm N}(q \, r_N) =\frac{3 j_1(q r_N)}{q r_N} e^{-q^2 s^2/2} \; ,
\end{equation}
where $r_N = 1.14 \, A^{1/3}$ fm~\cite{LEWIN199687} is the effective radius of the target nucleus, and $s_p=0.9$ fm~\cite{LEWIN199687} is the skin thickness of the nucleus.

\subsection{Combined Rate}
%%%%%%%%%%%%%%%%%%%%%%%%%%%%
The total rate involves integrating over $q$ from $0$ to the $q_{\rm max}$, set by Eq.~\eqref{eq:qmax}. Thus, the rate always involves the low $q$ region, and will involve the high $q$ region if $q_{\rm max} \gg 1/r_{\rm atom}$. There is also an intermediate regime we must address: when $q \sim 1/r_{\rm atom}$, i.e., the transition between where the charged particles see individual nucleons versus the full atom. The full rate can be written parametrically as:
\begin{equation}
   \left. \Gamma_{\rm tot}\right|_{\textsc{cpb}} = \int_0^{q_{\rm max}} \frac{d \left. \Gamma_{\rm tot}\right|_{\textsc{cpb}}}{dq} =  \int_0^{q_\text{low}} \frac{ \left. d\Gamma_{\rm tot}\right|_{\textsc{cpb}\text{,low q}}}{dq} + \int_{q_\text{low}}^{q_\text{high}} \frac{\left. d\Gamma_{\rm tot}\right|_{\textsc{cpb}\text{,inter q}}}{dq} +\int_{q_\text{high}}^{q_\text{max}} \frac{d \left. \Gamma_{\rm tot}\right|_{\textsc{cpb}\text{,high q}}}{dq}  \; .
\end{equation}
For a cosmic ray - proton interaction, the intermediate regime, $q \sim 1/r_{\rm atom}$, could be addressed by adding to the Coulomb potential the impact of the negative charge of the electrons screening the positive charge of the nucleus~\cite{Landau:1991wop}:
\begin{equation}
    \tilde U (\mathbf{q}) = \frac{ e}{4\pi q^2} \left( Z +  F(\mathbf{q}) \right) \quad \text{ where }\quad  F(\mathbf{q}) = \int d^3 \mathbf{r} \, n(\mathbf{r}) \, e^{i \mathbf{q}\mathbf{r}} \; ,
\end{equation}
where $n(\mathbf{r})$ is the density of the cloud of electrons. For $q \gg 1/r_{\rm atom}$, $F(q) \to 0$, due to the oscillatory behavior of the exponential. For $q \ll 1/r_{\rm atom}$, the coherent interactions of the charged particles with the dipole induced in the neutral atom dominate. Both low-q and high-q regimes can be well-separated from the intermediate regime by choosing $q_{\rm low} \ll 1/r_{\rm atom}$ and $q_{\rm high} \gg 1/r_{\rm atom}$, respectively. For the numbers shown in this section, we choose the threshold to be $q_{\rm low} = 1/(10\ r_{\rm atom})$ and $q_{\rm high} = 10/r_{\rm atom}$, where we take $r_{\rm atom}$ to be the Bohr radius, i.e. $r_{\rm atom} = 5 \times 10^{-11} \text{ m}$~\cite{ParticleDataGroup:2020ssz}. We summarize these regions in Fig.~\ref{fig:diagram}.
\begin{figure}[t]
\begin{equation*}
\begin{gathered}
\begin{tikzpicture}[line width=1.5 pt,node distance=1 cm and 1 cm]
\filldraw[ pattern=north west lines,draw=none] (1,-0.2) rectangle (5,0.2);
\draw[thick,->] (-3,0) -- (9,0);
\node(A) at (-1.5,0.4){$A_0(r) = 1/r^4$};
\node (AB) at (-1, -0.4) {low-$q$};
\node (B) at (1,0.5) {$q_{\rm low} = \frac{1}{10 \, r_{\rm atom}}$};
\node (Bo) at (1,0) {$|$};
\node (BC) at (3, -0.4) {intermediate regime};
\node (CD) at (7.5, 0.4) {$A_0(r) = 1/r$};
\node (C) at (5, 0.5) {$q_{\rm high} = \frac{10}{r_{\rm atom}}$};
\node (CDd) at (7.5, -0.4) {high-$q$};
\node (Co) at (5, 0) {$|$};
\node (q) at (9.5,0){$q$};
\end{tikzpicture}
\end{gathered}
\end{equation*}
\caption{Integration regimes in $q$ have been considered by employing different approaches.}
\label{fig:diagram}
\end{figure}
As we show in Table~\ref{tab:CRs}, the low-q and high-q contributions are already enough for the solar wind to erase the fringe for MAQRO, and to dominate over the QNL error for all of the other experiments. The contribution from the intermediate regime can only worsen the visibility, so we do not need to consider it to make the claim that the solar wind will be an important background for matter-wave interferometers in space.

We note that Ref.~\citep{Kunjummen:2022uzx} also calculated the effect of cosmic rays on atom interferometers. However, they only calculated the effect on the phase, and they ignored the effect of the solar wind. As in our case, they find that cosmic rays produce a negligible effect on atom interferometers.

\begin{table}[h]
\begin{tabular}{ |c || c  c c c|}
\hline
Solar wind & \ \ \ MAQRO \ \ \  &\ \ \  BECCAL\ \ \  &\ \ \  GDM\ \ \  & \ \ \ AEDGE\ \ \  \\
\hline
\hline
low-$q$ rate [$\text{s}^{-1}$] 
& $2.2 \times 10^1$ & $2.0 $ 
& $1.9 \times 10^2$ & $7.0 \times 10^3$ \\
\hline
high-$q$ rate [$\text{s}^{-1}$] & $4.4 \times 10^{3}$ & $5.5 \times 10^1$ & $5.5 \times 10^{3}$  & $5.5 \times 10^{5}$ \\
\hline
\end{tabular}

\vspace{0.3cm}

\begin{tabular}{ |c || c  c c c |}
\hline
Cosmic rays & MAQRO  & BECCAL & GDM  & AEDGE \\
\hline
\hline
low-$q$ rate [$\text{s}^{-1}$]                
& $1.7 \times 10^{-11}$                               
& $1.4 \times 10^{-12}$                              
& $1.4 \times 10^{-10}$                           
& $5.2\times 10^{-9}$  \\  
\hline
high-$q$ rate [$\text{s}^{-1}$] 
& $ \ 2.8 \times 10^{-9} \ $ 
& $ \ 3.5 \times 10^{-11} \ $ 
& $ \ 3.5 \times 10^{-9} \ $ 
& $3.5 \times 10^{-7}$ \\
\hline
\end{tabular}
\caption{Total decoherence rate $\Gamma_{\rm tot}$ in $s^{-1}$ caused by solar wind (upper table) and  by cosmic rays (lower table).} 
\label{tab:CRs}
\end{table}

\section{Other Backgrounds}\label{sec:other_bkgds}
In this section, we discuss the final two particle backgrounds that we consider in this paper: zodiacal dust and solar neutrinos. Unlike the previous two backgrounds, these particles produce negligible amounts of decoherence in the experiments that we consider. We include these calculations for completeness.

\subsection{Zodiacal Dust}\label{sec:dust}

Interplanetary dust in the inner solar system mostly comes from the fragmentation of comets and meteoroids. The mass (and size) distribution of zodiacal dust grains is inferred from the frequency of micro-craters on lunar samples and meteorites, as well as data from experiments such as Pioneers 8, 9, 10 and 11, Helios, and HEOS-II, amongst others \citep{Grun1985}. Interstellar dust also flows into the solar system as the Sun moves through the galaxy. However, we assume that this is subdominant at 1 AU \citep{Grun1997}. We take the interplanetary meteoroids model \citep{Grun1985}, which estimates the dust at 1 AU, for our zodiacal dust mass distribution. 

Since these dust grains have sizes $a \sim 10^{-4}  \, \mu {\rm m} - 1 \, {\rm cm}$ (masses $10^{-8} \, {\rm g} < m < 1 \, {\rm g}$) 
, which is comparable to the size of the clouds in the AIs, we use the geometric cross section to quantify the probability of interaction:
\begin{equation}
\sigma_{\rm geo} = \pi (r_{\rm cloud} + a)^2  \; ,
\end{equation}
where the radius of the dust grain is related to its mass by adopting a material density of $\rho_{\rm dust} \sim 3 \text{ g/cm}^3$~\cite{GieseGrun1976}.
The total rate between Zodiacal dust and the AI clouds is given by:
\begin{equation}\label{eq:ratedust}
\begin{split}
\left. \Gamma_{\rm tot}\right|_{\rm dust} &= \int dm  \, \frac{d\Phi_{\rm dust}}{dm} \sigma_{\rm geo} \, N_{\rm clouds} \; \\
& = \int \frac{dm}{m} \, \left(\frac{dn_{\rm dust}}{d\log m} \right) \frac{v_{\rm dust}}{k \, \ln 10} \, \pi \left(r_{\rm cloud} + \left(\frac{3m}{4\pi \rho_{\rm dust}}\right)^{1/3} \right)^2 N_{\rm clouds}\;,
\end{split}
\end{equation}
where $dn/d\log m$ is the spatial number density, which we take from~\cite{Grun1985}, $N_{\rm clouds} = 2$ is the number of clouds involved in the experiment, and $k$ is a constant (in particular, $k=4$ for an isotropic flux). Notice that such a contact interaction always resolves the two paths, thus $p_{\rm dec}=1$. As with all of our backgrounds, we assume the zodiacal dust is isotropic. Above, $v_{\rm dust} = v_0 \sqrt{r/r_0}$ is the orbital velocity of the dust particles, with $r$ referring to the radial distance from the Sun, $r_0 = 1$ AU, and $v_0 = 20$ km/s, taken from meteor observations and satellite measurements~\cite{Grun1985}. We take $r=1$ AU for the missions we consider. Table~\ref{tab:dust} shows the decoherence rate for the experiments we consider. Overall, these are negligible.  
\begin{table}[h]
\begin{tabular}{ |c || c | c| c| c| c |}
\hline
Dust & MAQRO &  BECCAL & GDM & AEDGE \\
\hline
Rate [$\text{s}^{-1}$] & $2.2 \times 10^{-14}$ & $2.4\times 10^{-8}$ & $1.1\times 10^{-6}$ & $9.6 \times 10^{-6}$  \\
\hline
\end{tabular}
\caption{Total decoherence rate $\Gamma_{\rm tot}$ in $s^{-1}$ caused by Zodiacal dust.} 
\label{tab:dust}
\end{table}

\subsection{Solar Neutrinos}\label{sec:neutrinos}
%%%%%%%%%%%%%%%%%%%%%%%%%%%%%%%%%%%%%%%%%%%%%%%%%
Solar neutrinos are byproducts of nuclear fusion processes in the core of the Sun. Their fluxes depend on the solar temperature, opacity, chemical composition, nuclear cross sections, and other solar features. Typically, solar models are employed to calculate the solar neutrino fluxes~\cite{Bahcall:1997eg,RevModPhys.60.297,Bahcall:2004pz}. Solar neutrinos are created throughout the ``pp-chain'' whereby the Sun fuses hydrogen ions into helium ions in a number of steps. The first reaction, which we call ``pp" here, is: $p+p\to d + e^+ + \nu_e$, $E_\nu \leq 0.42 \text{ MeV}$. This step composes 90\% of the solar neutrino flux \cite{Bahcall:1997eg}. Other components of the solar neutrino flux are the pep-neutrinos ($p + e + p \to d + \nu_e$, $E_\nu = 1.44 \text{ MeV}$ -monochromatic), $\phantom{I}^{7}\text{Be}$-neutrinos ($e + \!\!\!\phantom{I}^{7}\text{Be} \to \!\!\!\phantom{I}^7\text{Li}+\nu_e$, $E_\nu = 0.86 \text{ MeV (90\%)},\, 0.38 \text{ MeV (10\%)}$), $\!\phantom{I}^8B$-neutrinos ($\!\phantom{I}^8B \to \!\!\!\phantom{I}^8\text{Be}^*+e^++\nu_e$, $E_\nu < 15 \text{ MeV}$), and hep-neutrinos ($\!\phantom{I}^3\text{He}+p \to \!\!\!\phantom{I}^4 \text{He}+ e^+ +\nu_e$, $E_\nu < 18.8 \text{ MeV}$)~\cite{Bahcall:2004in}. The CNO (carbon-nitrogen-oxygen) cycle has a subdominant contribution (about $ 1\%$)~\cite{Bahcall:2004in}. In Fig.~\ref{fig:kinematics}, we show the total flux of solar neutrinos from different production channels. 

The spin-averaged modulus squared of the scattering amplitude for a neutrino interacting with a nucleon through the vector coupling of the $Z$ boson is given by:
\begin{equation}
    \begin{gathered}
\begin{tikzpicture}[line width=1.5 pt,node distance=1 cm and 1.5 cm]
\coordinate[label=left:$\nu_L$](nu1);
\coordinate[right = 0.75 cm of nu1](aux1);
\coordinate[right = 0.75 cm of aux1,label=right:$\nu_L$](nu2);
\coordinate[below = 0.3 cm of aux1](v1);
\coordinate[below=0.5cm of v1,label=right:$Z$](vZ);
\coordinate[below = 1.5 cm of aux1](aux2);
\coordinate[above = 0.3 cm of aux2](v2);
%\coordinate[below= 0.15 cm of v2,label=below:$g_{ZN}$](gZN);
\coordinate[above= 0.15 cm of v1,label=above:$g_{Z\nu}$](gZN);
\coordinate[left = 0.75 cm of aux2,label=left:${\rm N}$](N1);
\coordinate[right = 0.75 cm of aux2,label=right:${\rm N}$](N2);
\draw[fermionnoarrow](nu1)--(v1);
\draw[fermionnoarrow](v1)--(nu2);
\draw[vector](v1)--(v2);
\draw[fermion](N1)--(v2);
\draw[fermion](v2)--(N2);
\draw[fill=orange] (v1) circle (.07cm);
\draw[fill=orange] (v2) circle (.07cm);
\end{tikzpicture}
\end{gathered} \qquad 
\begin{gathered}
\overline{|{\cal  M}|^2} \simeq 8 |g_{Z{\rm nuc}}|^2 |g_{Z\nu V}|^2 \frac{2 (M^2-s)^2 + 2 s t +  t^2}{m_Z^4} \; ,
\end{gathered}
\end{equation}
where the target mass $M$ is the mass of the nucleon (N). 
In the laboratory frame and using that $q \ll M$ (see Fig.~\ref{fig:kinematics}), $s  = M^2 + 2 \omega M$, and $t = -q^2$. Hence: 
\begin{equation}
\overline{|{\cal M}|^2} \simeq 8 \frac{|g_{Z {\rm nuc}}|^2 |g_{Z\nu V}|^2 }{m_Z^4} \left(8 \omega^2 M^2 - 2 M q^2 (2 \omega + M)+q^4\right) \; .
\end{equation}
The vector coupling between the Standard Model $Z$ boson and a left-handed neutrino is given by $|g_{Z\nu V}| = e / 2 \sin2\theta_W$, where the electroweak angle is $\sin^2\theta_W = 0.23$~\cite{ParticleDataGroup:2022pth}.

Neutrinos from the sun are always below 20 MeV, and therefore their interactions with nuclei are within the coherent regime. Hence, the effective coupling between the Z boson and the nucleus is given by~\cite{PhysRevD.98.053004}
\begin{equation}
\begin{split}
|g_{Z{\rm nuc}}|^2 &= \frac{\alpha \, \pi}{ (\sin 2 \theta_W )^2}  |Z(4 \sin^2\theta_W -1)+(A-Z)|^2 |F_N(q r_N)|^2 \\
&\simeq \frac{\alpha \, \pi}{ (\sin 2 \theta_W )^2}  (A-Z)^2  \; ,
\end{split}
\end{equation}
where $A$ and $Z$ are the numbers of nucleons and protons in the nucleus of the target, respectively. In the last equality above, we have used that $\sin^2\theta_W \sim 1/4$ and that in the coherent regime $F_N(qr_N) \sim 1$, where $F_N(qr_N)$ is the nucleus form factor (see Eq.~\eqref{eq:FN}).

Furthermore, in matter interferometers, if the de-Broglie wavelength of the interaction is larger than the size of the cloud, {\it i.e.} $q < 1/r_{\rm cloud}$, there is another coherent enhancement across all of the atoms, and the probability amplitude scales as\footnote{Note that such an enhancement does not occur in experiments for direct detection of neutrinos based on nuclear recoil. This is because the lowest recoil energy these experiments are sensitive to ($E_{\rm rec} \sim 1$ meV) sets a lower threshold on the momentum transfer, i.e. $q_{\rm min} \gtrsim 1/r_{\rm atom}$.} $N_{\rm atoms}^2$.
Thus, the scattering amplitude reads~\cite{Drukier:1984vhf}:
\begin{equation}
\begin{split}
    \overline{|{\cal M}|^2} &= 8 G_F^2 M^2 \omega^2 \left(1 -  \frac{q^2 }{4\omega^2} \left(\frac{2 \omega}{M} + 1 \right)+\frac{q^4}{8 \omega^2 M^2}\right) \\
    & \quad \times  (A-Z)^2 \left(|F_N(q r_{\rm N})|^2 + N_{\rm atoms}  |F_{\rm AI}(q r_{\rm cloud})|^2\right) \; ,
    \end{split}
\end{equation}
where $G_F=4 \pi \alpha / (\sqrt{2}M_Z^2 \sin^2 2\theta_W )$ is the Fermi constant, and $F_{\rm AI}(q r_{\rm cloud})$ is given in Eq.~\eqref{eq:HelmetFF}. 
Converting $\overline{|{\cal M}|^2}$ to $|{\sf M}|^2$ (see Eq.~\eqref{eq:conversion} in Appendix~\ref{app:formalism}) and plugging this into Eq.~\eqref{eq:master}, we find:
\begin{eqnarray}
    \left. \Gamma_{\rm tot} \right|_{\nu_{\odot}}
    &&\simeq \frac{G_F^2}{4\pi} \int_0^\infty d\omega  \frac{d\Phi}{d\omega} \int_0^{2\omega} dq \, q   \left( 1 - \frac{q^2}{4\omega^2}\left(\frac{2\omega}{M}+1\right) + \frac{q^4}{8 \omega^2 M^2}\right) \\
    &&\qquad \qquad \times  N_{\rm atoms} (A-Z)^2 \left(
     |F_N(q r_N)|^2 + N_{\rm atoms} F_{\rm AI}(q r_{\rm cloud})|^2\right)\,\Big[ 1 - {\rm sinc} ( q \Delta x ) \Big] \; , \nonumber
\end{eqnarray}
where the nuclear contribution is neglected because the incoming energy of solar neutrinos is not high enough to resolve the nucleus of the atom, see Fig.~\ref{fig:kinematics}.
In Table~\ref{tab:solarnu}, we show the results for the decoherence rate due to solar neutrinos for the space-based matter-wave interferometers that we consider. As with the zodiacal dust case, these rates are negligible.
\begin{table}[h]
\centering
\begin{tabular}{ |c || c  c c  c  |}
\hline
 $\nu_\odot$ & MAQRO &  BECCAL & GDM &  AEDGE \\
\hline
Rate [$\text{s}^{-1}$] & $ \ 3.2 \times 10^{-24} \ $ & $ \ 4.1 \times 10^{-26} \ $ & $ \ 4.1 \times 10^{-24} \ $ & $ \ 4.1 \times 10^{-22} \ $   \\
\hline
\end{tabular}
\caption{Total decoherence rate $\Gamma_{\rm tot}$ in $s^{-1}$ caused by solar neutrino fluxes.} 
\label{tab:solarnu}
\end{table}

\section{Discussion \& Conclusions}\label{sec:conclusion}

In this paper, we calculated the decoherence from solar photons, the solar wind, galactic cosmic rays, zodiacal dust, and solar neutrinos on space-based matter-wave interferometer experimental concepts. We developed the formalism necessary to treat both relativistic and non-relativistic particles with non-uniform spectral density.  As summarized in Tab.~\ref{tab:full_res}, the solar wind and solar photons are important backgrounds to these proposed experiments. In addition, solar photons, along with cosmic rays, could erase the fringe of matter interferometers like MAQRO. 

Given the decoherence rates we find for solar photons and the solar wind, it is likely that these experiments will need to consider shielding carefully. While shielding against solar photons is relatively simple, shielding against the solar wind could be quite difficult and costly. Given some of the large cloud separations, especially for GDM, this could prove to be especially prohibitive. 

Another possible resolution to dealing with these backgrounds is to change the orbit of the spacecraft. In this work, we considered an Earth-like orbit. To mitigate against both solar photons and the solar wind, a spacecraft could be positioned in a relatively low Earth orbit. This would allow the spacecraft to be in Earth's shadow for part of its orbit and within the protective envelope of Earth's magnetic field. BECCAL would already be in such a position, given that it is proposed to be on the ISS. Solar orbits at larger semi-major axes and outside of the ecliptic plane could also be possible. However, a detailed study of the decoherence given different orbital configurations is left to future work.

Our work also assumes that these backgrounds are isotropic. However, all of these backgrounds are actually highly directional. Including the directional information would not have a large effect on the total rate -- a 50\% enhancement or reduction at most (see Appendix A of Ref.~\cite{AIDM}). Nevertheless, this directional information could help with separation from other signals, such as decoherence from dark matter particles \citep{AIDM}.

In this work, we solely considered the decoherence effects of these background particles. However, we expect these particles to also induce phase effects in these matter-wave interferometer experiments. The highly anisotropic nature of these backgrounds will likely mean that these phase shifts could be subtracted out of any science analysis, given sufficient knowledge of the spacecraft orbit and background flux directions. However, we leave a detailed analysis of this topic to future work.

Matter-wave interferometers are sensitive experiments and have the potential to be important probes of gravitational waves and dark matter. However, careful modeling and mitigation of backgrounds is needed for them to reach their potential. In this paper, we have shown that cosmic rays, in particular, will need to be shielded by all of the proposed experiments. In addition, we have developed the formalism needed to calculate rates for varying fluxes of these particles. The inclusion of shielding and orbit mitigation in space-based matter-wave interferometer designs is recommended.

\acknowledgements
The authors would like to thank Leonardo Badurina, Sheng-wey Chiow, Ryan Plestid and Jess Riedel for especially helpful discussions. We would also like to thank Haiming Deng, Gerard Fasel, Brandon Hensley, Ed Rhodes, Roger Ulrich, and Edoardo Vitagliano. 
KP was supported by the U.S. Department of Energy, Office of Science, Office of High Energy Physics, under Award No.~DE-SC0021431, the Quantum Information Science Enabled Discovery (QuantISED) for High Energy Physics (KA2401032).
This work is supported by the U.S. Department of Energy, Office of Science, Office of High Energy Physics, under Award Number DE-SC0011632 and by the Walter Burke Institute for Theoretical Physics. This work was performed in part at Aspen Center for Physics, which is supported by National Science Foundation grant PHY-2210452.

\appendix

\section{Cross Section Formalism}
\label{app:formalism}
%%%%%%%%%%%%%%%%%%%%%%%%%%%%%%%%%
The number of events contributing to the decoherence can also be obtained from the scattering cross section of the background particle with the matter-wave interferometer cloud, as opposed to using the formalism based on Fermi's Golden rule, described in Section~\ref{sec:formalism}. The cross section is defined as the number of events per effective area ($A_T$) per incoming ($N_{\rm back}$) and target  ($N_{\rm target}$) particles: $\sigma = \#_{\rm events} \, A_T / (N_{\rm back} \, N_{\rm target})$.
The rate of events causing decoherence can then be written as a function of the decoherence cross section and the flux of incoming particles ($\Phi = N_{\rm back} / ( A_T \times  \text{time})$ ),  
\begin{equation}\label{eq:rateXsec}
    \frac{d \Gamma_{\rm tot}}{dt} = N_{\rm target} \int d\omega \left(\frac{1}{4\pi} \frac{d\Phi}{d\omega}\right) \int d \Omega \int_0^{q_{\rm max}(\omega)} \,  dq \,  \frac{d \sigma}{dq}.
\end{equation}
Above we have assumed that the flux of incoming particle is isotropic. The decoherence form factor, $p_{\rm dec}(\mathbf{q}\cdot \mathbf{\Delta x})$, is included in the cross section. In the laboratory frame, with $\hat{\mathbf{z}}$ axis aligned with the momentum of the incoming particle $\mathbf{p}$ (see Section~\ref{sec:formalism}), the momentum transfer can be written as:
\begin{equation}
    \mathbf{q} = (q \sin \theta_{\mathbf{q}\mathbf{p}} \cos \phi_{\mathbf{q}\mathbf{p}}, q \sin \theta_{\mathbf{q}\mathbf{p}} \sin \phi_{\mathbf{q}\mathbf{p}}, q \cos \theta_{\mathbf{q}\mathbf{p}}) \; ,
\end{equation}
where $q \equiv |\mathbf{q}|$. In this frame, the differential cross-section is given by:
\begin{equation}
\begin{split}
 \frac{d\sigma}{dq}& = \frac{1}{2\omega 2 \omega' |v - v'|} \frac{q^2}{(2\pi)^2}   \\
 &  \times \int d \cos \theta_{\mathbf{q}\mathbf{p}} \frac{\overline{|{\cal M}|^2}}{2 E_i 2 E_f}  \left( 1 + N_{\rm target} |F_{\rm AI}(q)|^2\right)\, \delta^{(0)}(\omega + \omega' - E_i - E_f)\times \int d \phi_{\mathbf{q} \mathbf{p}} \left( 1 - e^{i \mathbf{q} \cdot \mathbf{\Delta x}}\right) \; ,
 \end{split}
\end{equation}
where the relative velocity between the incoming particle ($v$) and the target ($v'$) can be expressed as $\omega \, \omega' |v-v'| = M \sqrt{\omega^2 - m^2}$. The term $\left( 1 + N_{\rm target} |F_{\rm AI}(q)|^2\right)$ accounts for the potential Born enhancement due to elastic scattering with the target and the type of experiment (see Eq.~\eqref{eq:HelmetFF}). Above, ${\cal M}$ refers to the scattering matrix element with the standard normalization, \textit{i.e.}, quantum states normalized with their energy. We have explicitly written the form of $p_{\rm dec}(\mathbf{q}\cdot \mathbf{\Delta x})$. Furthermore, we have exploited the fact that the matrix element in the lab frame does not depend on the azimuthal angle, $\phi_{\mathbf{q}\mathbf{p}}$, since the Mandelstam variables in the laboratory frame only depend on $\omega$, $q$, and the incoming particle and target masses. In other words:
\begin{equation}
    s = (p + k)^2 = M^2 + m^2 + 2 \omega M, \qquad \text{and} \qquad t = (k-k')^2 = 2M^2 \left( 1-  \sqrt{1+\frac{q^2}{M^2}}\right) \; .
\end{equation}
The decoherence form factor, however, does depend on $\phi_{\mathbf{q}\mathbf{p}}$. The vector parameterizing the separation of the two clouds, written in the laboratory frame, is:
\begin{equation}
   \mathbf{\Delta x} = (\Delta x \, \sin \theta_{\mathbf{\Delta x} \mathbf{p}} \cos \phi_{\mathbf{\Delta x}\mathbf{p}},\Delta x \, \sin \theta_{\mathbf{\Delta x}\mathbf{p}} \sin \phi_{\mathbf{\Delta x}\mathbf{p}} , \Delta x \, \cos \theta_{\mathbf{\Delta x}\mathbf{p}}) \;,
\end{equation}
where $\Delta x \equiv |\mathbf{\Delta x}|$, and $\theta_{\mathbf{\Delta x p}}$ is the angle between $\mathbf{\Delta x}$ and $\mathbf{p}$. Therefore, 
\begin{equation}\label{eq:J0int}
    \begin{split}
        &\int_0^{2\pi} d \phi_{\mathbf{q}\mathbf{p}}\left( 1 - e^{i q \Delta x (\sin \theta_{\mathbf{q}\mathbf{p}} \sin \theta_{\mathbf{\Delta x}\mathbf{p}} \cos (\phi_{\mathbf{q}\mathbf{p}}-\phi_{\mathbf{\Delta x}\mathbf{p}}) + \cos \theta_{\mathbf{q}\mathbf{p}}\cos \theta_{\mathbf{\Delta x}\mathbf{p}}}\right)\\
        &\qquad = (2\pi) \left( 1 - J_0(q \Delta x \, \sin \theta_{\mathbf{q}\mathbf{p}} \sin \theta_{\mathbf{\Delta x}\mathbf{p
        }}) e^{i q \Delta x \, \cos \theta_{\mathbf{q}\mathbf{p}} \cos \theta_{\mathbf{\Delta x}\mathbf{p}}}\right)\,.
    \end{split}
\end{equation}
 Since the angle between the incoming particle and the transferred momentum is a function of the incoming energy, the two variables can be exchanged according to the Jacobian:
\begin{equation}
    d \cos \theta_{\mathbf{q}\mathbf{p}} = - \frac{\omega}{q\sqrt{\omega^2 - m^2} } d \omega \; ,
\end{equation}
and use the energy delta function to perform the integral over $\omega$, leading to the following expression:\footnote{The integration of Eq.~\eqref{eq:J0int} over $d\Omega$ gives $8 \pi^2 (1 - \text{sinc} (q \Delta x))$. Note that the result is independent of $\theta_{\mathbf{q} \mathbf{p}}$, which becomes a function of $q$ after employing the energy delta function. See Appendix~\ref{app:ang} for an explicit derivation.} 
\begin{equation}
  \int d\Omega \,  \frac{d\sigma}{dq} = \frac{q}{8 M (\omega^2 - m^2)}\frac{\overline{|{\cal M}|^2} }{\sqrt{M^2 + q^2}} \left( 1 + N_{\rm target} |F_{\rm AI}(q)|^2\right) \left(1 - \frac{\sin (q \Delta x)}{q\Delta x}\right) \; ,
\end{equation}
where we have commuted the integrals over $q$ and $\Omega$, since the limits of $q$ do not depend on the angular phase space of the incoming particles (see the explicit form of $q_{\rm max}$ in Eq.~\eqref{eq:qmax}). 
Plugging the differential cross section into Eq.~\eqref{eq:rateXsec}, we obtain the master equation for the decoherence rate: 
\begin{equation}
    \frac{d \Gamma_{\rm tot}}{dt} =\frac{N_{\rm target}}{32\pi M } \int d \omega \frac{d\Phi}{d\omega}\frac{1}{\omega^2 - m^2} \int_0^{q_{\rm max}(\omega)} dq \, q\frac{\overline{|{\cal M}(\omega,q)|^2} }{\sqrt{M^2 + q^2}} \left( 1 + N_{\rm target} |F_{\rm AI}(q)|^2\right) \left(1 - \frac{\sin (q \Delta x)}{q\Delta x}\right) \; .
\end{equation}
From the equation above one can reproduce Eq.~\eqref{eq:master} in Section~\ref{app:formalism} by unity normalizing the states in the matrix element: 
\begin{equation}\label{eq:conversion}
    \overline{|{\cal M}|^2} = (2 \omega ) (2\omega') (2 E_i )(2 E_f ) \overline{|{\sf M}|^2}= 16 \omega \left( M +\omega - \sqrt{M^2+q^2}\right) M \sqrt{M^2+q^2} \ \ \overline{|{\sf M}|^2} \; ,
\end{equation}
and writing the spectral flux as the spectral density, which is related by the velocity of the incoming particle $v = \sqrt{1-\gamma^{-2}} = p / \omega$ (with $\gamma = \sqrt{1+p^2/m^2}$): 
\begin{equation}
    \frac{d\Phi}{d\omega} = \frac{dn}{d\omega} v = \frac{dn}{d\omega}\frac{\sqrt{\omega^2 - m^2}}{\omega} \; .
\end{equation}
Finally, the term $N_{\rm target} (1 + N_{\rm target} |F_{\rm AI}(q)|^2)$ can be identified as $S(q)$ (see Eq.~\eqref{eq:master} in Section~\ref{sec:formalism}).

\section{The Angular Phase Space Integration}

\label{app:ang}

In this section, we show how to reduce the six-dimensional phase space integration $d\omega d\Omega d^3 \mathbf{q}$ by integrating out the angular phase space. We first derive a general formalism applicable to any spectral density of the incoming particle with directional dependence, where we reduce the phase space integration to $d\omega d\Omega d q$. Then we consider the special case of an isotropic incoming source, which further reduces the integration to be two-dimensional, over $d\omega dq$.

For convenience, let us define:
\begin{equation}
   f(\omega, \Omega) \equiv \omega \frac{dn}{d\omega d\Omega}  \; .
\end{equation}
Rewriting the master equation in case of the (angular) phase space of the incoming source and the momentum transfer:
\begin{equation}
    \begin{split}
\Gamma_{\rm tot}
&=\frac{1}{(2 \pi)^2} \int \frac{d\omega}{\omega} \int dq \, q^2 \overline{|{\sf M}|^2}  S(q)
\int d\Omega_{\mathbf{p}}
 d\Omega_{\mathbf{q}}
 f(\omega, \Omega_{\mathbf{p}},t)
 \left(1 - e^{i {\bf q} \cdot {\bf \Delta x}} \right)
 \delta (E_f - E_i - \omega_{\bf q}) \; ,
    \end{split}
\end{equation}
where we specifically label the solid angle with the corresponding momentum.
Notice that, for all of the angles, the energy deposition, $\omega_{\mathbf{q}}$, only depends on the angle between $\mathbf{p}$ and $\mathbf{q}$, and that:
\begin{equation}
    \begin{split}
\frac{d \omega_{\mathbf{q}}}{d\cos\alpha_{\mathbf{p}\mathbf{q}}} = q \frac{\sqrt{\omega^2 - m^2}}{\omega - E_{\rm rec}} \; ,
    \end{split}
\end{equation}
where $E_{\rm rec} = \sqrt{M^2+q^2} - M$. Let us define $\alpha_0$ as $\left. E_{\rm rec} = \omega_{\mathbf{q}} \right|_{\alpha_{\mathbf{p}\mathbf{q}} = \alpha_0}$. Then we can solve for:
\begin{equation}
    \begin{split}
    \cos\alpha_0 = \frac{1}{2 \sqrt{\omega^2 - m^2} q} \left( q^2 + 2 \omega E_{\rm rec}  - E_{\rm rec}^2 \right) \; .
    \end{split}
\end{equation}
Accordingly, the Kronecker delta function enforcing energy conservation can be written as a function of the angle $\alpha_{\mathbf{p}\mathbf{q}}$:
\begin{equation}
    \begin{split}
 \delta (E_f - E_i - \omega_{\bf q}) =  \frac{\omega - E_{\rm rec}}{q \sqrt{\omega^2 - m^2}}
 \delta (\cos\alpha_{\mathbf{p}\mathbf{q}} - \cos\alpha_0) \; .
    \end{split}
\end{equation}
The angular part of the integration can be done generically:
\begin{equation}
    \begin{split}
& \int d\Omega_{\mathbf{p}}
 d\Omega_{\mathbf{q}}
 f(\omega, \Omega_{\mathbf{p}},t)
 \left(1 - e^{i {\bf q} \cdot {\bf \Delta x}} \right)
 \delta (\cos\alpha_{\mathbf{p}\mathbf{q}} - \cos\alpha_0)   \\
 = & ( 2 \pi ) \int d\Omega_{\mathbf{p}}f(\omega, \Omega_{\mathbf{p}},t)
 \Big[ 1 - J_0 ( q \Delta x  \sin\alpha_{0}\sin\theta_{\mathbf{p}\mathbf{\Delta x}}  ) e^{ i q \Delta x
\cos\alpha_{0} \cos\theta_{\mathbf{p}\mathbf{\Delta x}}  } \Big] \; .    \\
    \end{split}
\end{equation}
In deriving the above, we have used changes of variables, \textit{e.g.},
\begin{equation}
    \begin{split}
d \theta_{\mathbf{p}} d \phi_{\mathbf{p}}  = |J|\, d \alpha_{\mathbf{p}\mathbf{q}} d \gamma_{\mathbf{p}\mathbf{q}} \; ,
    \end{split}
\end{equation}
with
\begin{equation}
    \begin{split}
\cos\alpha_{\mathbf{p}\mathbf{q}} &= \cos \theta_{\mathbf{p}} \cos \theta_{\mathbf{q}} + \sin \theta_{\mathbf{p}} \sin \theta_{\mathbf{q}} \cos(\phi_{\mathbf{q}} - \phi_{\mathbf{p}}) \\
 \sin \gamma_{\mathbf{p}\mathbf{q}}  &=  \frac{\sin  \theta_{\mathbf{q}} \sin(  \phi_{\mathbf{q}} -  \phi_{\mathbf{p}}) }{\sin\alpha_{\mathbf{p}\mathbf{q}}} \; .
    \end{split}
\end{equation}
The Jacobian of the transformation is $|J| = \sin \alpha_{\mathbf{p}\mathbf{q}} /\sin  \theta_{\mathbf{p}}$. 
The phase space integration is now reduced to be over $d\omega d\Omega_{\mathbf{p}}dq$, independent of the angular phase space of $\mathbf{q}$.

In the case of an isotropic source, the density spectrum
does not depend on the solid angle, \textit{i.e.},~ 
\begin{equation}
    \begin{split}
    f(\omega, \Omega_{\mathbf{p}},t) = \omega \frac{d n (\omega,\Omega_{\mathbf{p}})}{d \omega d\Omega_{\mathbf{p}}}  \to  \frac{ \omega}{4 \pi} \frac{d n (\omega)}{d \omega} \; .
  \end{split}
\end{equation}
The angular part of the integration simplifies to be:
\begin{equation}
    \begin{split}
  &   ( 2 \pi ) \int d\Omega_{\mathbf{p}}f(\omega, \Omega_{\mathbf{p}},t)
 \Big[ 1 - J_0 ( q \Delta x  \sin\alpha_{0}\sin\theta_{\mathbf{p}\mathbf{\Delta x}}  ) e^{ i q \Delta x
\cos\alpha_{0} \cos\theta_{\mathbf{p}\mathbf{\Delta x}}  } \Big] \\
= &\frac{\omega}{2} \frac{d n}{d\omega} \int
d\phi_{\mathbf{p}\mathbf{\Delta x}}
d\cos\theta_{\mathbf{p}\mathbf{\Delta x}}
 \Big[ 1 - J_0 ( q \Delta x  \sin\alpha_{0}\sin\theta_{\mathbf{p}\mathbf{\Delta x}}  ) e^{ i q \Delta x
\cos\alpha_{0} \cos\theta_{\mathbf{p}\mathbf{\Delta x}}  } \Big] \\
= &( 2 \pi ) \omega \frac{d n}{d\omega}
 \Big[ 1 - {\rm sinc} ( q \Delta x ) \Big] \; .
    \end{split}
\end{equation}
Now the phase space integration has been reduced to be two-dimensional, $d\omega dq$, as has been used in the main text.

\bibliography{ref}

\end{document}